\documentclass[12pt,preprint]{aastex}

\newcommand{\dmm}{\mbox{$\Delta$m$_{15}(B)$}}

\lefthead{K. Krisciunas {\em et al.}}
\righthead{Photometry of Type Ia Supernovae}

\begin{document} 

\title{Optical and Infrared Photometry of the Type Ia 
Supernovae 1991T, 1991bg, 1999ek, 2001bt, 2001cn, 2001cz, and 2002bo}

\author{Kevin Krisciunas,$^{1,2,3}$
%
%
Nicholas B. Suntzeff,$^2$
Mark M. Phillips,$^1$ 
Pablo Candia,$^2$
Jose Luis Prieto,$^2$ 
Roberto Antezana,$^4$
Robin Chassagne,$^5$
Hsiao-Wen Chen,$^6$
Mark Dickinson,$^7$
Peter R. Eisenhardt,$^8$
Juan Espinoza,$^2$ 
Peter M. Garnavich,$^3$
David Gonzalez,$^2$
Thomas E. Harrison,$^9$
Mario Hamuy,$^1$
Valentin D. Ivanov,$^{10}$
Wojtek Krzeminski,$^1$
Craig Kulesa,$^{11}$
Patrick McCarthy,$^{12}$
Amaya Moro-Martin,$^{11}$
Cesar Muena,$^1$
Alberto Noriega-Crespo,$^{13}$
S. E. Persson,$^{12}$
Philip A. Pinto,$^{11}$
Miguel Roth,$^1$
Eric P. Rubenstein,$^{14}$
S. Adam Stanford,$^{15}$
Guy S. Stringfellow,$^{16}$
Abner Zapata,$^{17}$ 
Alain Porter,$^{7,18}$ and
Marina Wischnjewsky$^{4,18}$  }

\affil{$^1$Las Campanas Observatory, Carnegie Observatories, Casilla 601, La Serena, Chile \\
$^2$Cerro Tololo Inter-American Observatory, National Optical Astronomy \\ 
Observatory,\altaffilmark{19} Casilla 603, La Serena, Chile \\
$^3$University of Notre Dame, Department of Physics, 225 Nieuwland Science Hall, Notre 
Dame, IN 46556-5670 \\
$^4$Departmento de Astronom\'{i}a, Universidad de Chile, Casilla 36-D, Santiago, Chile \\
$^5$Ste. Clotilde Observatory, 58C2 Route Gabriel Mace, R\'{e}union Island \\
$^6$Center for Space Research, Massachusetts Institute of Technology, \\
77 Masachusetts Avenue, 37-664B, Cambridge, MA 02139-4307 \\
$^7$Kitt Peak National Observatory, National Optical Astronomy \\
Observatory,\altaffilmark{19} 950 N. Cherry Ave., Tucson, AZ 85719-4933 \\
$^8$Jet Propulsion Laboratory, MS 169-327, 4800 Oak Grove Drive, Pasadena, CA 91109 \\
$^9$Department of Astronomy, New Mexico State University, 1320 Frenger Mall, \\
Las Cruces, NM 88003-8001 \\
$^{10}$European Southern Observatory, Casilla 19001, Santiago, Chile \\
$^{11}$Department of Astronomy and Steward Observatory, 933 North Cherry Avenue, Tucson,
AZ 85721-0065 \\
$^{12}$Observatories of the Carnegie Institution of Washington, 813 Santa Barbara St.,
Pasadena, CA 91101-1292 \\
$^{13}$Spitzer Science Center, Caltech, MS 220-6, Pasadena, CA 91125 \\
$^{14}$Advanced Fuel Research, Inc., 87 Church St., East Hartford, CT 06108-3728 \\
$^{15}$Department of Physics, University of California at Davis, 1 Shields Avenue, \\
Davis, CA 95616-8677 \\
$^{16}$Center for Astrophysics \& Space Astronomy, University of Colorado, Box 389, \\
Boulder, CO 80309-0389 \\
$^{17}$Universidad de Concepci\'{o}n, Departamento de F\'{i}sica, Grupo de Astronom\'{i}a, 
Casilla 4009, Concepci\'{o}n, Chile \\
$^{18}$Deceased }
\altaffiltext{19}{The National Optical Astronomy
Observatory is operated by the Association of Universities for
Research in Astronomy, Inc., under cooperative agreement with the
National Science Foundation.}

\email {kkrisciu, pgarnavi@nd.edu \\
nsuntzeff, med@noao.edu \\
mmp, mario, miguel, wojtek@lco.cl  \\
jprieto@ctio.noao.edu \\
pcandia, juan, dgonzalez@ctiosz.ctio.noao.edu \\
robin.chassagne@wanadoo.fr \\
hchen@space.mit.edu \\
prme@kromos.jpl.nasa.gov \\
tharriso@nmsu.edu \\
vivanov@eso.org \\
pmc2, persson@ociw.edu \\
amaya, ppinto@as.arizona.edu \\
alberto@ipac.caltech.edu \\
ericr@afrinc.com \\
adam@igpp.ucllnl.org \\
guy.stringfellow@casa.colorado.edu \\
abzapata@udec.cl }

\begin{abstract}


We present optical and/or infrared photometry of the Type Ia supernovae
\objectname{SN~1991T}, \objectname{SN~1991bg},
\objectname{SN~1999ek}, \objectname{SN~2001bt}, \objectname{SN~2001cn},
\objectname{SN~2001cz}, and \objectname{SN~2002bo}.  All but one of
these supernovae have decline rate parameters \dmm\ close to the median 
value of 1.1 for the whole class of Type Ia supernovae.  The addition of
these supernovae to the relationship between the near-infrared absolute
magnitudes and \dmm\ strengthens the previous relationships we have
found, in that the maximum light absolute magnitudes are essentially
independent of the decline rate parameter.  (SN~1991bg, the prototype
of the subclass of fast declining Type Ia supernovae, is a special case.)
The dispersion in the
Hubble diagram in $JHK$ is only $\sim$0.15 mag.  The near-infrared
properties of Type Ia supernovae continue to be excellent measures of the
luminosity distances to the supernova host galaxies, due to the need
for only small corrections from the epoch of observation to maximum
light, low dispersion in absolute magnitudes at maximum light, and the
minimal reddening effects in the near-infrared.

\end{abstract}

\parindent = 0 mm

\keywords{supernovae, individual (SN~1991bg, SN~1999ek, SN~2001bt, SN~2001cn,
SN~2001cz, SN~2002bo) $-$ techniques: photometric}

\section{Introduction}

\parindent = 9 mm

Since the discovery of a precise relationship between the absolute
magnitude at maximum light and decline rate \citep{Phi93}, Type Ia
supernova light curves have become the most accurate method to measure
luminosity distances to objects at cosmological distances.  The
standard measure of the decline rate in the notation of Phillips,
\dmm, is the number of magnitudes that a Type Ia supernova declines in
the 15 days after the time of $B$-band maximum.  \dmm\ is now one of
many methods used to calibrate absolute magnitudes of Type Ia SNe
(supernovae). These include a modification of the \dmm\ method to
incorporate $BVI$ data by fitting discrete templates
\citep{Ham_etal96,Phi_etal99,Pri_etal05}; a similar multi-color
technique using a continuous set of multi-color templates with the
important additions of a self-calibrated error model
\citep{Rie_etal96,Jha02}; and a method that parameterizes the $BV$
light curves by ``stretching'' the time axis to fit any given light
curve set to a single template set \citep{Per_etal97}. Finally,
\citet{Wan_etal03} have shown that useful information 
such as reddening can be derived from plots of the filter by filter
magnitudes vs. the photometric {\em colors} (instead of vs. time), 
using data from the month after maximum light. All these
techniques rely on fitting optical multi-color light curves to a
template set which includes a single parameter which is independently
correlated to the intrinsic luminosity of the SN.

Until recently Type Ia SNe were primarily observed through the $BV$
and $RI_{KC}$ (Kron-Cousins) bands.  \citet{Jha02} has added the $U$
filter to the light curve sets and presents $UBVRI$ data for 44
objects.  The $U$-band data are particularly important for cosmology,
because many high-redshift SNe are measured in the $R$- and $I$-bands,
which may correspond to rest frame $U$-band light curves depending on
the redshift. The $U$-band light curves are sensitive to the
the metallicity of the progenitor if the $U$-band light
curve is measured before maximum light \citep{Hoe_etal98,Len_etal00}.

The compilation of near-IR (infrared) light curves of Type Ia SNe has
been slow due to a lag in IR detector technology with respect to
optical CCD detectors and the lack of telescopes dedicated to IR
observations.  The Caltech infrared group
\citep{Eli_etal81,Eli_etal85} presented the first extensive
observations of Type Ia SNe in the near infrared.  A complete summary
of IR observations obtained through 1998 is given by \citet{Mei00}.

In their Fig.~6 \citet{Eli_etal85} show an $H$-band Hubble diagram and
comment that Type Ia SNe at maximum light may be good standard
candles.  \citet{Mei00} analyzed the IR data available four years ago
and used Cepheid determinations of the distances to eight galaxies
which have hosted Type Ia SNe, concluding that the dispersion in absolute
magnitude 14 days after T($B_{max}$) was $\approx$0.15 mag over the
decline rate range 0.87 $\leq$ \dmm\ $\leq$ 1.31.
\citet{Kri_etal04a} presented the first Hubble
diagrams for Type Ia SNe based on $JHK$ magnitudes at maximum light.
Their data indicate that there are no obvious decline-rate relations
in the near-IR if we consider the objects with \dmm\ $\lesssim$ 1.7.
Thus, rather than standardizable candles, Type Ia SNe {\it at maximum
light} are excellent standard candles. The dispersions in the absolute
magnitudes at maximum light are roughly $\pm$ 0.15 mag for the near-IR
$JHK$ bands.  The near-IR data have the extra advantage that
extinction by interstellar dust is up to an order of magnitude less at IR
wavelengths compared to optical bands \citep{Car_etal89}.  Even for
objects significantly reddened at optical wavelengths, the extinction
corrections are small in the IR, thus eliminating a previously serious
potential source of random and systematic error in the distance
determinations.

There are other advantages to using the near-IR magnitudes in
conjunction with the optical ones.  In our Papers I through V we
discussed the use of the $V$ {\em minus} near-IR loci for Type Ia
SNe. Except in the unusual case of SN~2000cx (Paper IV) we found
that the unreddened loci of $V-J$ or $V-H$ allowed the
determination of A$_V$ with smaller random errors and less
worry about systematic errors. This is because: 1)
unreddened Type Ia SNe appear to exhibit predictable $V$ {\em
minus} near-IR color curves, allowing us to derive sensible color
excesses for objects which are reddened; and 2) the total
extinction A$_V$ is equal to these color excesses to a factor of
about 1.3. Thus we can use, say, $V-H$, as a method of measuring
A$_V$ without introducing a significant uncertainty due to a
non-standard extinction law.  Since R$_V$ can be as low as 1.8
(see Paper I for a discussion of SN~1999cl), serious systematic
errors of distance can result if one uses only optical data and
assumes R$_V$ = 3.1 to convert $E(B-V)$ into A$_V$.

A final advantage to the use of colors such as $V-J$ or $V-H$ can be
made by appealing to our experience with measuring stellar effective
temperatures \citep{Bla_etal90,Bes_etal98}. A color based on two
filters close in wavelength such as $B-V$ is a very sensitive function
of $T_{eff}$ but also the metallicity of the photosphere. A color such
as $V-K$ is also sensitive to metallicity, but because of the very
large wavelength difference between the bands, the effect of
temperature is much larger making the metallicity effect secondary. We
are far from understanding the continuum formation of complicated
non-hydrogen NLTE photosphere of a Type Ia SN, but the observed range
in color and ionization seen at maximum light can be explained by a
simple temperature sequence where the more luminous SNe are hotter
\citep{Nug_etal95}. If this were true, by using a color such as $V-H$,
which may be a less complicated function of T$_{eff}$ rather than
$B-V$, we may be able to uncover a clearer relationship between the
observed colors and the underlying physical temperature and
luminosity.

Since 1999 we have been engaged in regular monitoring of bright Type
Ia SNe (typically $V_{max} < 15.5$), at Cerro Tololo Inter-American
Observatory (CTIO) and Las Campanas Observatory (LCO) with the goal
to construct a near-IR Hubble diagram with a sample of 30 Type Ia SNe.
We are motivated by the success of the Cal\'{a}n/Tololo SN survey
\citep{Ham_etal93b} which published a similar number of optical light
curves of Type Ia SNe.  Given the importance of high redshift SNe for
cosmology, i.e. evidence for a non-zero cosmological constant
\citep{Rie_etal98,Per_etal99,Ton_etal03,Kno_etal03}, a well-calibrated
sample of nearby SNe studied in the near-infrared will allow us to
better understand the dereddened properties of the population of Type
Ia SNe, and ultimately allow us to better anchor the distant sample to the 
nearby one.

This is the sixth in a series of data papers on optical and infrared
photometry of Type Ia supernovae (SNe).  We have already
presented data on \objectname{SN~1999aa}, \objectname{SN~1999cl}, and
\objectname{SN~1999cp} \citep[Paper I]{Kri_etal00};
\objectname{SN~1999da}, \objectname{SN~1999dk}, \objectname{SN~1999gp},
\objectname{SN~2000bk}, and \objectname{SN~2000ce} \citep[Paper
II]{Kri_etal01}; \objectname{SN~2001el} \citep[Paper III]{Kri_etal03};
\objectname{SN~2000cx} \citep[Paper IV]{Can_etal03}; and
\objectname{SN~1999ee}, \objectname{SN~2000bh}, \objectname{SN~2000ca},
and \objectname{SN~2001ba} \citep[Paper V]{Kri_etal04b}.  \citet{Jha02}
has presented other optical data for \objectname{SN~1999aa},
\objectname{SN~1999cl}, \objectname{SN~1999gp}, \objectname{SN~2000ce},
and \objectname{SN~2000cx}.  Extensive optical photometry of
\objectname{SN~1999ee} has been presented by
\citet{Str_etal02}. \citet{Li_etal01b} presented data for
\objectname{SN~2000cx}.

Here we present optical and/or near-IR photometry of the Type Ia
supernovae \objectname{SN~1991T}, \objectname{SN~1991bg},
\objectname{SN~1999ek}, \objectname{SN~2001bt},
\objectname{SN~2001cn}, \objectname{SN~2001cz}, and
\objectname{SN~2002bo}.  Three of these objects were discovered by one
of us (RC) as part of an ongoing SN search with a 0.3-m f/2.7
telescope sited at R\'{e}union Island (latitude = S 21\arcdeg
06$^{\prime}$, longitude = E 55\arcdeg 36$^{\prime}$).  Except for
SN~1991bg, the group of
supernovae presented in this paper has a \dmm\ decline rate close to
the median value for all Type Ia SNe. In that sense, this is a
``normal'' group of Type Ia supernovae which should be representative
of the average population.

\section{Observations}

\subsection{Photometric Calibration}

We used \citet{Lan92} standards for the calibration of our optical
photometry. The field star magnitudes were derived from aperture
photometry in the {\sc iraf} environment.\footnote[20] {{\sc iraf} is
distributed by the National Optical Astronomy Observatory, which is
operated by the Association of Universities for Research in Astronomy,
Inc., under cooperative agreement with the National Science
Foundation.}  Typically, we derive the photometry of the SNe
themselves using point spread function ({\sc psf}) magnitudes using
{\sc DAOPHOT} \citep{Ste87,Ste90}.  When necessary, we first performed
image subtraction using templates taken after the SNe have faded away.
Some of our subtractions were carried out with a script based on Andrew
Phillips' {\sc iraf} package {\sc psfmatch}, while others were carried
out using a script written by Brian Schmidt and based on the algorithm
of \citet{Ala_Lup98}.

Except for the case of SN~1991bg, we
calibrated the IR magnitudes of the field stars near the SNe using
the near-IR standards of \citet{Per_etal98} and observations from the
LCO 1-m telescope.  We have found that the
Two Micron All Sky Survey (2MASS) magnitudes of field stars brighter
than 15th magnitude are, on average, 0.03 mag brighter than the
derived $J_s$, $H$, and $K_s$ photometry on the system of
\citeauthor{Per_etal98} With this offset we were able to assess the
quality of the photometry of the field stars if they were observed by
us on few photometric nights.

The photometry of the local standards is given in Tables 1 and 2.  The
finding charts are shown in Figures \ref{find_chart}a through \ref{find_chart}e.

Photometry of a given SN, obtained with multiple telescopes, shows
systematic differences from telescope to telescope.  Often this
amounts to several hundredths of a magnitude, but once a SN begins to
enter the nebular phase two to three weeks after maximum light, these
systematic differences can amount to several tenths of a magnitude.
In Papers III, IV, and V we used a method of spectroscopically-based
filter corrections to correct the photometry for the non-stellar
spectral energy distributions of the SNe and the differences of the
filter transmission curves of the filters in different instruments.
The problem of bringing the photometry of Type Ia SNe onto a standard
system was discussed by \citet{Sun00}, and a correction - the
``S-correction'' method - was given in \citet{Str_etal02}; the reader
should consult this paper for the definition of the linear transformations.
We have found, for example, comparing YALO and CTIO 0.9-m $B$-band
photometry of Type Ia SNe, that starting some 30 days after the time of 
$B$-band maximum there is a systematic difference of 0.07 mag owing to
the filter differences.  In the $V$-band the systematic differences
are 0.05 mag, but with the opposite arithmetic sense.  Without taking
this into account YALO photometry gives $B-V$ colors 0.12 mag different
in the tail of the color curve.  This clearly makes a significant
difference if one derives a $B-V$ color excess from photometry
at this epoch.  As found by \citet{Kri_etal03} in the case of SN~2001el,
these systematic differences can be largely eliminated by correcting
the photometry to the system described by \citet{Bes90}.

The applications of S-corrections to date have only reduced the
systematic errors of the photometry taken on different telescopes, not
eliminated them. \citet{Eli03} suggests that these theoretical
corrections are never as good as one would hope due to the difficulty
of measuring the instrument/filter/detector system efficiencies and
that the real S-corrections should be calculated empirically as the
observed differences of the photometry obtained with different telescopes and
instruments. However, there are very few SNe which have been well
sampled with multiple telescopes, so we cannot yet set up appropriate
look up tables.

In Figs. \ref{scorr_bv} and \ref{scorr_ri} we show optical filter
corrections to the \citet{Bes90} system, while in Fig. \ref{scorr_jhk}
we show IR corrections from the YALO filters to the system of
\citet{Per_etal98}. The trends in these figures are not precise, and it is
very important to get spectral atlases of other SNe to see if the
S-corrections of other SNe approximate what we find in the cases of
SNe 1999ee, 2001el, and 2002bo.

The filters and CCDs used in the CTIO 1.5-m CCD camera are nominally
the same as those in the 0.9-m camera.  We have assumed that the
S-corrections derived for the $UBVRI$ filters are the same for the two
cameras.

Our principal goal with the photometry of the five supernovae 
of mid-range decline rates discussed
in this paper is to derive color excesses (i.e. E(B$-$V), giving
extinction corrections) and infrared magnitudes at maximum light.  The
photometry {\em presented in the tables} includes the color corrections
based on observations of standard stars.  For further analysis we have
applied the S-corrections and K-corrections to all our data where
possible, the one exception being the Steward IR data for SN~1999ek
where we do not have a good prescription for the system throughputs.  
In the case of SN~2002bo we explicitly list the S-corrections (Tables 12
and 13). The IR K-corrections are based on spectra of SN~1999ee and are
given in Paper V.

We have used a new program to carry out the $\Delta$m$_{15}$ analysis
\citep{Pri_etal05}, which is a variant on the method given in
\cite{Phi_etal99}.  The $BVRI$ light curves of a given SN are fitted
simultaneously, minimizing a $\chi^2$ statistic constructed as the
difference between the data and the model (i.e. templates in each filter
plus magnitudes at maximum), weighted by the uncertainties in the data
and model. The parameters of the fit (unknown variables) are: the time
of $B$-band maximum, the magnitudes at maximum in each filter ($BVRI$),
the value of $\Delta$m$_{15}$, and the total reddening
E($B-V$)$_{total}$.\footnote[21]{Effectively, the code
gives total reddening.  It assumes that the Galactic reddening from
\citet{Sch_etal98} is correct.  \citet{Arc_Goo99} discuss possible
systematic errors in the Galactic reddening of Schlegel et al.  But
if the Schlegel et al. reddening used by us is systematically too {\em large} by a 
certain amount, then our derived value of the host galaxy reddening should
be systematically too {\em small} by exactly the same amount.  This is to say
that there should be no systematic error in our effective value of
total reddening, which is what we desire most for accurate supernova
distances.}  The host reddening is simply the total reddening
minus the Galactic reddening given by \citet{Sch_etal98}. The new \dmm\ 
code uses the same 6 templates based on 7 objects that were used by
\citet{Phi93}, \citet{Ham_etal96}, and \citet{Phi_etal99}, 
along with the well-sampled  light
curves of 8 Type Ia SNe observed more recently.  The code adopts
K-corrections similar to those described by \citet{Ham_etal93a}. In
contrast to previous versions of $\Delta$m$_{15}$ analysis, the Prieto
code allows for a continuous variation of $\Delta$m$_{15}$. The analysis
produces a covariance matrix, allowing us to calculate the 1-$\sigma$
statistical uncertainties of the parameters.

\subsection{SN 1991T}

SN 1991T is the prototype of the slowly declining Type Ia SNe.
\citet{Phi_etal99} give \dmm\ = 0.94 $\pm$ 0.05 for this object.  While
SN~1991T was originally considered significantly overluminous
\citep{Rie_etal96}, \citet{Gib_Ste01} recently obtained a Cepheid
distance to the host galaxy, NGC 4527. Their distance modulus of $m-M$ =
30.562 $\pm$ 0.085 mag indicates that this object has optical absolute 
magnitudes at maximum {\em comparable} to Type Ia SNe with mid-range 
decline rates.

Data of \citet{Lir_etal98} indicate that the $B$-band maximum of
SN~1991T occurred on 1991 April 29.2 UT (JD 2,448,375.7).  
\citet{Men_Car91} reported $JHK$ data taken on April 17.96 UT, some 11.2
days prior to T($B_{max}$).  Menzies (2004, private communication)
indicates that these measurements were taken with a 0.75-m telescope
at South African Astronomical Observatory and a single channel photometer
giving a 36 arcsec diameter beam.  The instrument looked at two sky
patches approximately 3 arcmin north and south of the SN location.

\citet{Har_Str91} have previously reported $JHK$ photometry of
SN~1991T obtained on 1991 June 22 UT with the Australian National
University's 2.3-m telescope at Siding Spring.  Here we report two
previously unpublished nights of ANU 2.3-m infrared data (see Table 3), taken
5.05 and 6.10 days after T($B_{max}$).  These observers used a
single channel InSb detector, 10 arcsec diameter beam, and a 20 arcsec 
chop in the north-south direction.  Finally, we note that these data are 
on the MSSSO infrared standard system.  For subsequent calculations we
have slightly corrected them to the system of \citet{Eli_etal82}
using the linear transformations given by \citet{McG94}.

Given that these previously unpublished data were taken within 10 days
of the presumed time of the IR maxima, we can then use the templates of
Paper V and the Cepheid distance of the host to estimate the absolute
magnitudes at maximum. We adopted E($B-V$)$_{Gal}$ = 0.022 mag from
\citet{Sch_etal98} and E($B-V$)$_{host}$ = 0.14 $\pm$ 0.05
\citep{Phi_etal99}.  With R$_V$ = 3.1, A$_V$ = 0.50 mag.  Standard
\citet{Car_etal89} extinction parameters then give A$_J$ = 0.14, A$_H$ = 
0.10, and A$_K$ = 0.06 mag.

\subsection{SN 1991bg}

SN 1991bg is the prototype of the rare fast declining Type Ia SNe.
According to \citet{Li_etal01a}, only 16 percent of Type Ia SNe 
discovered in a distance-limited survey are like SN~1991bg.  The 
fast decliners are intrinsically red at maximum
light compared to mid-range and slow decliners and lack the secondary
maximum in the $I$-band light curves.  They are subluminous by 2 to 3
magnitudes in the $V$- and $B$-bands.  The fast decliners are probably
explosions of C-O white dwarfs yielding the smallest amounts ($\approx$
0.1 M$_{\sun}$) of $^{56}$Ni \citep[][and references therein]{Str_Lei04}.

\citet{Fil_etal92} and \citet{Lei_etal93} give optical CCD
photometry of SN~1991bg.  The maximum brightness in $V$ occurred on
1991 December 14.7 UT (JD2,448,605.2).  The time of $B$-band maximum is
not as well contrained by the fewer $B$-band data, but if SN~1991bg
behaved like SN~1999by, an equally fast decliner \citep{Gar_etal04},
then the $B$-band maximum of SN~1991bg probably occurred
about 2 days prior to T($V_{max}$). We will assume that T($B_{max}$)
occurred at JD2448603.2.  For the purpose of calculating absolute
magnitudes we assume a distance modulus of $m-M$ = 31.32 $\pm$ 0.11
\citep{Ton_etal01} for NGC 4374 (the host of SN~1991bg), which is 
based on the method of Surface Brightness Fluctuations (SBF's).

\citet{Por_etal92} presented an AAS poster discussing five nights  
of $JHK$ data and the bolometric light curve of SN~1991bg.  However,
Alain Porter died in 1993 and the data were never published.  We have
acquired the raw data used by Porter for the critical 3 nights near
maximum light taken by \citet{Sta_etal95}.  The
data were taken with the Kitt Peak National Observatory 1.3-m
telescope using the Simultaneous Quad-color Infrared Imaging Device  
(SQIID).  This camera contained four low quantum efficiency 256
$\times$ 256 PtSi arrays.  The plate scale on the 1.3-m telescope was
1.36 arcsec per pixel.  SQIID took simultaneous $JHKL^{\prime}$
images, but the $L^{\prime}$ images were not saved.

We used images obtained by Stanford et al. to characterize the    
non-linearities of SQIID.  We calibrated SN~1991bg directly using     
bright IR standards of \citet{Eli_etal82} on the photometric night of
15 December 1991 and determined the $JHK$ magnitudes of field stars A
and J of \citet{Lei_etal93}.  We confirmed that no color terms need
to be used for the determination of the $J$ and $K$ data with SQIID,
but there is a substantial color term for $H$, namely $-$0.261 $\pm$
0.053, scaling $H-K$.  This is in agreement with the value found by
Stanford et al. (see their \S3).

We list the $JHK$ magnitudes of
stars A and J in Table 2.  In Table 3 we give the $JHK$ photometry of 
SN~1991bg derived from SQIID images.  The two remaining nights of the
Porter data in the $J$-band were obtained by E. Lada et al.  (2004   
private communication) in January of 1992 but are not included in our
paper.  \citet{Por_etal92} state that SN~1991bg faded by three
magnitudes in the $J$-band in the month after maximum light.

\subsection{SN 1999ek}

SN 1999ek was discovered in UGC 3329 by \citet{Joh_Li99} from images
taken on 1999 October 20.5 (JD 2,451,472.0) and 21.4 UT at $\alpha$ =
05:36:31.6, $\delta$ = +16:38:18 (equinox 2000), at 11.7 arcsec west
and 12.1 arcsec south of the core of UGC 3329.  \cite{Str_etal99}
reported that spectra taken on October 25 and 26 UT indicated that
SN~1999ek was a Type Ia SN at or near maximum light.
\cite{Jha_etal99} report that a spectrum taken on October 30.46 UT
confirms the SN typing and age with respect to maximum.

This supernova, located at a Galactic latitude of $-$8 degrees, has
a large Galactic reddening according to \citet{Sch_etal98} at E($B-V$)
= 0.561 mag.  We note that \citet{Arc_Goo99} find that for A$_V \; >$
0.5 mag, the \citeauthor{Sch_etal98} maps overestimate the reddening
by a factor of 1.3 to 1.5.

All our optical data of SN~1999ek were obtained with the CTIO 0.9-m
telescope. Our image subtraction templates were obtained with the CTIO
1.5-m telescope on 2001 November 12 UT.  The $JH$ infrared templates
were obtained with the LCO 1-m telescope on 2000 November 17 UT.

Our IR data of this object were obtained with the 1.2-m telescope of
the Fred L.  Whipple Observatory at Mt. Hopkins, Arizona (1 night),
the YALO 1-m at CTIO (7 nights) and the Steward Observatory 2.3-m at
Kitt Peak (5 nights).  The Steward data were obtained with a 256
$\times$ 256 HgCdTe {\sc nicmos} detector \citep{Rie_etal93}. We found
that the images could not be flattened and sky-subtracted well.
As a result, the Steward data should be considered only approximate
($\pm$ 0.08 mag).  In Table 4 we give the internal errors of the
photometry.

In Tables 4 and 5 we give the optical and IR data of SN~1999ek. We
took no $K_s$-band imagery of this field with the LCO 1-m, and the one
night of FLWO imagery was not photometric.  We have opted to use the
K-band values of the nearby field stars from 2MASS to calibrate the single
night of K-band data of SN~1999ek.

In Fig. \ref{99ek_all} we show the optical and IR light curves of
SN~1999ek.  We include some optical photometry of \citet{Jha02} which
is consistent with our data at the 0.01 mag level at JD
2,451,487. 

Using the light curve fitting code of \citet{Pri_etal05}
we derive E($B-V$)$_{total}$ = 0.76 $\pm$ 0.05 mag for SN~1999ek.  The
implied host reddening is E($B-V$)$_{host}$ = 0.76 $-$ 0.56 = 0.20
mag.  The implied total extinction is A$_V$ = 2.37 $\pm$ 0.16 mag,
assuming R$_V$ = 3.1 for both components of reddening.

Using the unreddened $V-H$ locus for mid-range decliners (Paper V,
Table 14), we derive E($V-H$) = 1.83 $\pm$ 0.05 mag for SN~1999ek
using only the LCO infrared data.  This gives A$_V$ = 2.25 $\pm$ 0.12
mag.  Thus, in the case of SN~1999ek, the method of combining optical
and IR photometry for mid-range decliners discussed in Papers I, II,
III, and V gives a value of A$_V$ consistent with the $\Delta$m$_{15}$
method.

\subsection{SN 2001bt}

SN 2001bt was discovered by \citet{Cha01a} from images taken on 2001
May 24.0 (= JD 2,452,053.0) and 24.9 UT.  It is located at $\alpha$ =
19:13:46.8, $\delta$ = $-$59:17:23 (equinox 2000), 14.1 arcsec west
and 17.1 arcsec north of the nucleus of IC 4830.  From a spectrum
obtained on May 26 UT with the LCO 2.5-m telescope \cite{Phi_Kri01}
deduced that SN~2001bt was a Type Ia SN at or near maximum.

Tables 6 and 7 contain our photometry of
SN~2001bt itself.  Fig. \ref{01bt_all} shows the optical and IR
photometry of SN~2001bt.

The data obtained with the LCO 2.5-m telescope have much smaller
internal errors than the data from the YALO 1-m telescope at a
comparable epoch near maximum light.  Our values of the maximum
IR magnitudes of SN~2001bt hinge much more on the LCO photometry,
for which no filter corrections are necessary as well, to place
the data on the filter system of \cite{Per_etal98}.

\subsection{SN 2001cn}

SN 2001cn was also discovered by \citet{Cha01b} from images of 2001
June 11.95 (= JD 2,452,072.45) and 12.85 UT.  It is located at
$\alpha$ = 18:46:17.8, $\delta$ = $-$65:45:42 (equinox 2000), 2.6
arcsec west and 17.9 arcsec south of the nucleus of IC
4758. \citet{Can_etal01} report that a spectrum obtained by
D. Norman and K. Olsen on June 14.3 UT with the CTIO 1.5-m telescope
showed that SN~2001cn was a Type Ia SN near maximum light.

Tables 8 and 9 contain our photometry of
SN~2001cn itself.  Fig. \ref{01cn_all} shows the optical and IR
photometry of SN~2001cn.

\subsection{SN 2001cz}

Yet another SN was discovered by \citet{Cha01c} from images taken on
2001 July 4.64 (= JD 2,452,095.14) and 5.61 UT.  SN~2001cz is located
at $\alpha$ = 12:47:30.2, $\delta$ = $-$39:34:48 (equinox 2000), 0.6
arcsec west and 31.4 arcsec south of the nucleus of NGC
4679. \citet{Pas_etal01} obtained a spectrum on July 12.01 UT with the
Danish 1.54-m telescope at La Silla, finding the new object to be a
Type Ia SN within 2 d of maximum light.  Tables 10 and 11 contain our
photometry of SN~2001cz itself. Fig. \ref{01cz_all} shows the optical
and IR photometry of SN~2001cz.

In Fig. \ref{find_chart}d we mark a previously unknown
RR Lyrae star.  Its $V$ magnitude ranges from 16.20 to 16.72, and
its period is 0.6405 d.  For more details see \citet{Kri_etal04c}.
According to \citet{Sch_etal98}, the Galactic reddening in the direction
of NGC 4679 is E($B-V$) = 0.092.  Assuming R$_V$ = 3.1, the $V$-band
extinction is then 0.285 mag.  With $\langle$V$\rangle$ = $\frac{1}{2}$ (V$_{max}$
+ V$_{min}$) + 0.07 \citep[p. \ 27]{Smi95}, and adopting a mean absolute magnitude
M$_V$ = +0.7 \citep{Lay_etal96}, the distance to the new RR Lyr star is roughly
12.9 kpc.  It is located some 5 kpc above the Galactic plane.

\subsection{SN 2002bo}

SN 2002bo was independently discovered by \citet{Cac_Hir02} on 2002
March 9 UT.  \citet{Nak_etal02} give a position of $\alpha$ =
10:18:06.5, $\delta$ = +21:49:42 (equinox 2000), some 11.6 arcsec east
and 14.2 arcsec south of the nucleus of the SA galaxy NGC 3190.
\cite{Kaw_etal02} and \citet{Ben_etal02} obtained spectra on March 9
and 10, respectively, which showed that SN~2002bo was a Type Ia SN 10
to 12 days before maximum.  According to NED, the heliocentric
redshift of NGC 3190 is 1271 km s$^{-1}$, or $z$ = 0.0042, a very nearby
galaxy.

NGC 3190 is a member of the Leo III group
\citep{Gar93}. \citet[their Table 1]{Ton_etal01} determined SBF distances
of two possible members of this group.  They obtained $m-M$ = 32.66
$\pm$ 0.18 for NGC 3193 and $m-M$ = 31.86 $\pm$ 0.24 for NGC 3226.
From dynamical information \citet{Ben_etal04} obtained a distance
modulus for SN~2002bo of 31.67 mag on a scale of H$_0$ = 65 km
s$^{-1}$ Mpc$^{-1}$. Correcting to an H$_0$ = 72 scale, this becomes
$m-M$ = 31.89, close to the measured SBF distance modulus of NGC 3226,
but in significant disagreement with the distance modulus of NGC 3193.

A similar distance estimate can be obtained by using the flow model of
\cite{Ton_etal00}.  One uses the Supergalactic coordinates of NGC~3190
to calculate the flow velocities along the vector towards the
galaxy. A distance is recovered when the flow velocity matches the
radial velocity of the galaxy with respect to the Cosmic Microwave
Background.  For SN~2002bo we find $v_{flow}$ = $v_{CMB}$ at a
distance modulus of 31.69 on the H$_0$ = 78.4 scale of
\citeauthor{Ton_etal00}. This is equivalent to a distance modulus of
31.87 on an H$_0$ = 72 scale, corresponding almost exactly to the SBF
distance modulus of NGC~3226 on the same scale. We will adopt the
distance modulus for NGC 3226, 31.86 $\pm$ 0.24, as the distance
modulus to SN~2002bo.  The equivalent recession velocity in the CMB
frame would be 1696 km s$^{-1}$.

For the CTIO 0.9-m data we used images of 2003 January 3 UT as image
subtraction templates.  For the CTIO 1.5-m data we used subtraction
templates obtained with that telescope on 2003 January 30 UT.  All the
YALO 1-m data are derived from {\sc psf} magnitudes without image
subtraction templates.

Without S-corrections the $U$-band photometry of SN~2002bo obtained
with the YALO, CTIO 0.9-m and 1.5-m telescopes shows systematic
differences up to 0.4 mag due to the lack of ultraviolet response
below 3500\AA\ in the SITe CCDs on the CTIO telescopes.  If we
calculate S-corrections based on our best knowledge of the $U$-band
filters for the CTIO and the YALO telescopes, we still find small
differences between the two filter systems. To improve the fits, we
shifted the two filter functions until we matched the color terms from
synthetic photometry of spectrophotometric standards with color terms
obtained from actual broad-band photometry of stars. We found we
needed to shift the nominal CTIO 0.9-m $U$-band filter profile 40
\AA\ to shorter wavelengths.  The {\sc andicam} $U$-band filter
profile had to be shifted 30 \AA\ toward longer wavelengths. The color
term, based on actual photometry of stars, as opposed to synthetic
photometry, is +0.107 $\pm$ 0.010 (scaling $U-B$) for the CTIO 0.9-m
telescope and $-$0.084 $\pm$ 0.020 for YALO.

\citet{Ben_etal04} give MJD = 52356.0 $\pm$ 0.5 $\equiv$ Julian Date
2,452,356.5 as the time of T($B_{max}$) for
SN~2002bo.  From CTIO data only we obtain
T($B_{max}$) = JD 2,452,356.5 $\pm$ 0.2, in agreement with the value
of \citeauthor{Ben_etal04}.

In Tables 12 and 13 we present our optical and IR photometry of
SN~2002bo.  In Fig. \ref{02bo_all} we show the optical and IR light
curves of SN~2002bo, including the data of \citet{Ben_etal04}.
\citet{Sza_etal03} have also published 11 nights of $VRI$ data
for SN~2002bo.

The premaximum data of SN~2002bo do not fit our $JHK$ templates very
well, the data being roughly 0.3 mag fainter than the template (see
Fig. \ref{jhk_new}).  However, it should be noted that 
our templates are based on very few data
points obtained more than a week before T($B_{max}$).  
Still, because we actually measured this SN at maximum, we do not
need to use our templates to determine what the observed maxima were.

In Fig. \ref{02bo_vir} we show the $V$ {\em minus} near-IR color
curves of SN~2002bo.  We also plot the unreddened loci for mid-range
decliners (Papers I and V), adjusted by the implied color excesses to
minimize the reduced $\chi^2$ values of the fits.  The reader will
notice immediately that the loci based on other objects do not fit
these color curves very well, especially around day 10.  The reduced
$\chi^2$ values range from 2 to 4, and the data fail to match the
templates by up to 0.3 mag at 10 days after T($B_{max}$).

The implied values of A$_V$ are 0.71 $\pm$ 0.16 from E($V-J$),
0.64 $\pm$ 0.09 from E($V-H$), and 0.51 $\pm$ 0.08 from E($V-K$).  
While these three values are statistically equal, within
the errors (the weighted mean being 0.59 $\pm$ 0.06), the color
excesses listed in Fig. \ref{02bo_vir} do not {\em increase} from
$V-J$ through $V-K$, as they ought to.  The optical
photometry of SN~2002bo indicates that E($B-V$)$_{total}$ = 0.39
$\pm$ 0.02, implying A$_V$ = 1.21 {\em if} R$_V$ = 3.1.  
Conversely, if the host galaxy extinction is A$_{V,host}$ = 0.59
$-$ 3.1 $\times$ 0.025 = 0.51 [using E($B-V$)$_{Gal}$ from Table
15], and E($B-V$)$_{host}$ = 0.365, it follows that R$_{V,host}$ =
1.40.  This would imply very non-standard dust in the host
galaxy. We feel it is most likely that the odd implied color
excesses of SN~2002bo are due to its having abnormal unreddened
colors.

From spectroscopic evidence \citet{Ben_etal04} suggest that the
unusually low temperature of SN~2002bo, the presence of
high-velocity intermediate-mass elements and the low abundance of
carbon at early times indicates that the silicon burning in this
object penetrated to much higher layers than in more normal Type
Ia SNe.  Whatever the physical explanation, the spectral energy
distributions (SEDs) in the $V$- and near-IR bands were
considerably different than other Type Ia SNe we have studied.  
As a result there may be {\em no} color index that can be used to
obtain a reliable value of A$_V$ for this unusual object.

\subsection{Spectral Observations}

Spectral observations were obtained for three of the program
supernovae. A log of the observations is given in Table 14. For the
spectra taken with the CTIO 1.5-m telescope, the facility Cassegrain
spectrograph was used with two grating setups. The gratings have the
following specifications listed as (lines per mm, blaze, resolution at
5000 \AA, \AA ngstr\"{o}ms per pixel): grating 09, (300, 4000 \AA, 7.3 \AA,
2.88 \AA\ px$^{-1}$); grating 13, (150, 5000 \AA, 14 \AA, 5.7 \AA\
px$^{-1}$).  The CCD is a Loral 1200 $\times$ 800 pixel detector with
excellent ultraviolet response. The detector fringes redward of 8000
\AA. Although the wavelength coverage using grating 13 covers more than
an octave, we did not use an order blocking filter to remove the blue
contamination from first order red. These spectra will be contaminated
by blue light longward of 6600 \AA.  Because of flexure in the
spectrograph, wavelength calibration spectra were taken at every slew
position. We observed spectrophotometric standards from the lists of
\citet{Ham_etal92,Ham_etal94}. The supernovae were observed through 2
and 10 arcsec slits, and the standards through a 10 arcsec slit. The
slit length was roughly 4 arcmin.

The data were reduced to fluxes with the {\sc specred} package in
{\sc iraf}. The CCD data were bias subtracted, trimmed, and flat-fielded
using either dome flats or twilight spectra. In the case of the
twilight spectra, the solar lines were removed by filtering. The data
were wavelength calibrated to an accuracy of $\approx$0.1 pixel residual in
the fit. The underlying galaxy spectra was removed by interpolating
the galaxy flux across the position of the supernova. The spectra were
brought to an absolute flux scale with reference to the Hamuy et al.
spectrophotometric standards and standard extinction tables.  To cover
the full ultraviolet range, the table of fluxes of the standard stars
was extrapolated to 3100 \AA\ by fitting a quadratic function over the
range 3300 to 3650 \AA. The small slit data were corrected to the flux scale
of the 10 arcsec observations. Not all the nights were photometric, so the
flux scale should be considered as a relative scale.

We present the spectra in Fig. \ref{spectra}. We also include two spectra of
SN~1999ee which were observed on the same nights as SN~1999ek but not
included in the \citet{Ham_etal02} study of that supernova. We also
include two earlier time spectra of SN~1999ee from \citet{Ham_etal02}
to provide a spectral sequence of a normal Type Ia over the epoch of
observations of our program supernovae. From a comparison of the
features of the SNe shown in Fig. \ref{spectra} with spectra given by 
\citet{Fil97}, we find that the program
supernovae were caught within a week of T($B_{max}$).

\section{Discussion}

Using the infrared light curve templates of Paper V, which indicate
IR maxima about 3 days prior to T($B_{max}$), we find that
SN~1991T, the prototype of the slowly declining Type Ia supernovae,
had implied IR absolute magnitudes at maximum quite comparable to
our sample of slow decliners and mid-range decliners.  The IR data
given by \citet{Men_Car91}, obtained 11.2 days prior to
T($B_{max}$), are between 1.1 and 1.8 magnitudes brighter than our
templates.  Either their measurements included a significant amount
of host galaxy light, or the behavior of this object long before
T($B_{max}$) was different than any object studied to date.

While the observations of SN~1991bg presented here and those of
SN~1999by by \citet{Hoe_etal02} and \citet{Gar_etal04} do not
allow us to constrain fully the IR maxima of these two objects,
because of the rarity of the fast declining Type Ia SNe and
the paucity of IR data on these objects, some comments are
in order.  Firstly, SN~1999by was observed close
enough to maximum light that we can derive $H$- and $K$-band 
maxima for SN~1999by.  The $J$-band maximum of SN~1999by was
likely no more than 0.1 mag brighter than the observation
made 2 days prior to T($B_{max}$).  Likewise, the $JHK$ observations
of SN~1991bg 0.85 days after T($V_{max}$) must have been less
than 0.1 mag from the IR maxima.  Thus, we can compare the
luminosities and intrinsic colors of two fast decliners (with essentially 
zero host reddening) with the larger number of Type Ia SNe of
mid-range and slow decline rates.  The case of
SN~1986G, which had \dmm\ almost as large, is complicated by its 
substantial and uncertain reddening \citep[see][and references 
therein]{Kri_etal04a}.

In Fig. \ref{vjhkmax} we show the colors at maximum of the better
sampled objects shown in Fig. 12 of \citet{Kri_etal04b} along
with the near-maximum colors of SNe 1991bg and 1999by.  The purpose
of this graph is merely to demonstrate the intrinsic redness of the fast
decliners in the V {\em minus} near-IR color indices.  Fig. 16
of \citet{Gar_etal04} shows the $B-V$ colors at maximum vs.
\dmm\ and also demonstrates that Type Ia SNe with \dmm\ $\gtrsim$ 1.7
exhibit much redder colors than Type Ia SNe with lesser decline
rates.

In Table 15 we summarize the basic data for the five SNe with mid-range 
decline rates presented in
this paper. Listed are: the velocity of recession with respect to the
Cosmic Microwave Background radiation (using the NED velocity
calculator for objects with redshift greater than 3000 km s$^{-1}$),
the time of $B$-band maximum, the decline rate \dmm, and the Galactic
and host contributions of the $B-V$ color excess.  We assumed values
of E($B-V$)$_{Gal}$ from \citet{Sch_etal98}.  In Table 15 the values
of E($B-V$)$_{host}$ are derived from optical data alone.

In Table 16 we give the
apparent magnitudes at maximum for the objects discussed in this paper
and their extinction-corrected values.  We adopt the values of
A$_{\lambda}$ / A$_V$ = 0.282, 0.190, and 0.114 for the $J$-, $H$-,
and $K$-bands, respectively, given by \citet{Car_etal89} in their Table 3, 
column [5].

In Fig. \ref{jhk_hubble} we show the Hubble diagrams of Type Ia SNe
for the near-IR $JHK$ bands, adding objects from this paper to Fig. 2
of \citet{Kri_etal04a}.  In Fig. \ref{absmag_z} we show the
$JHK$ absolute magnitudes at maximum light versus the redshift
for Type Ia SNe of mid-range and slow decline rates.  There is
no obvious trend, indicating that the absolute magnitudes are uniform
to a look back time of 0.5 Gyr.  Fig. \ref{absmag_dm15} shows the
$JHK$ absolute magnitudes at maximum light versus the decline rate
parameter \dmm.  As found by \citet{Kri_etal04a}, the absolute
magnitudes show no obvious correlations with the optical decline rate
parameter until we consider the fastest decliners.
Thus, Type Ia SNe with \dmm $\lesssim$ 1.7
can be considered standard candles at maximum light in the near-IR.
We note that in spite of the peculiarities of the light curves and
implied color excesses of SN~2002bo, the small {\em infrared} extinction
corrections and the uniformity of IR absolute magnitudes at maximum give
this object IR absolute magnitudes similar to ``normal'' Type Ia SNe.

In Table 17 we give the mean absolute magnitudes of our growing sample
of Type Ia SNe and the 1-$\sigma$ distribution widths.  This is an
update of the corresponding table of \citet{Kri_etal04a}.  None
of the numbers has changed significantly.

Not only is the extinction much lower in the near-IR compared to the
rest frame optical bands, but we can use the $JHK$ infrared templates
to derive the maximum magnitudes if data are taken in the time frame of
$-$12 to +10 days with respect to T($B_{max}$). If the implied
distances of high redshift SNe derived from {\em rest frame} $J$- or
$H$-band photometry are in agreement with the distances derived from
rest frame $UBV$ photometry, that will increase our confidence in the
adopted extinction corrections.  Also, it will give us confidence that
Type Ia SNe that exploded billions of years ago are fundamentally the
same type of objects we now observe in nearby galaxies.

Because Type Ia SNe are standard candles in the near-IR, except for
the objects with the largest possible decline rates, it would be
ideal if high-redshift supernovae are observed in the $H-$ or $K-$bands
so that we might exploit the advantages of observing these important
objects at least in the rest frame $J$-band. The IR bandpasses are
closer to notch filters\footnote[22]{Having filter profiles with flat
tops and minimal wings.} than the optical bands, and we would expect
that the K-corrections should be small at ``magic'' redshifts where the
$J$ band redshifts to the effective wavelength of $H$ or $K$, or the
$H$ band redshifts to $K$. This happens at z=0.33 ($J \Rightarrow H$,
$H_{max}=21.8$), z=0.73 ($J \Rightarrow K$, $K_{max}=23.1$), and z=0.30
($H \Rightarrow K$, $K_{max}=21.8$) for the LCO system of
\citet{Per_etal98}. It is unlikely that ground-based telescopes can
reach the $K$-band detection at z=0.73, but the $H$-band detection of
S/N=15 at z=0.3 is roughly one hour on an 8-m telescope in good
conditions. With a few supernovae sampled roughly 4 times near maximum
light, the effects of acceleration could be easily seen in the
rest-frame $J$-band magnitudes.

\section{Conclusions}

In this paper we have provided optical and/or infrared photometry of the
Type Ia supernovae \objectname{SN~1991T}, \objectname{SN~1991bg},
\objectname{SN~1999ek}, \objectname{SN~2001bt},
\objectname{SN~2001cn}, \objectname{SN~2001cz}, and
\objectname{SN~2002bo}.  

The previously unpublished near-IR data of SN~1991T are important
because SN~1991T is the prototype of the slowly declining Type Ia
SNe. These data were obtained close enough to T($B_{max}$) that they
allow us to estimate the absolute magnitudes at maximum of this
object.  Just as \citet{Gib_Ste01} found regarding the optical
absolute magnitudes at maximum, we find that SN~1991T had IR
luminosities at maximum comparable to other, spectroscopically
normal, objects.

Also, because of the importance of SN~1991bg as the prototype of
the fast declining Type Ia SNe, we felt it was worthwhile to
re-reduce and publish the few infrared data near the epoch of
maximum light without further delay.  We also do this to honor our
late colleague Alain Porter. Within the errors, SN~1991bg had a
decline rate equal to that of SN~1999by.  The absolute magnitudes
near maximum light in the IR of these two objects were essentially
the same, and roughly half a magnitude fainter than Type Ia SNe of
mid-range and slow decline rates.

\citet{Ben_etal04} have shown that SN~2002bo
exhibited unusual SEDs.  As a result, we cannot confidently derive a
value of the extinction (A$_V$) toward this object, because we cannot
compare its colors to other objects with similar SEDs which are known
to be unreddened.  Four objects presented in this paper have
optical and IR light curves which are like many ``normal'' Type Ia
SNe.  

We have used a new program to carry out the $\Delta$m$_{15}$ analysis
for the $BVRI$ data \citep{Pri_etal05}.  As one can see in the
corresponding figures, the derived light curve templates fit the
$BVRI$ data very well.

Following up the work of \citet{Kri_etal04a,Kri_etal04b}, we derived
the extinction-corrected apparent (and absolute) magnitudes of
five new mid-range decliners
in the near-IR and confirmed that there is no apparent
dependence of the absolute magnitudes at maximum versus the decline rate
or the redshift for Type Ia SNe with \dmm\ $\lesssim$ 1.7.
The rms scatter of the absolute magnitudes is
roughly $\pm$~0.15 mag in the $JHK$ bands.  Thus, an individual object
can give a distance good to $\pm$~7 percent.

\acknowledgements  

We made use of the NASA/IPAC Extragalactic Database (NED), the {\sc
simbad} database, operated at CDS, Strasbourg, France, and the Two
Micron All-Sky Survey (2MASS). We thank the Space Telescope Science
Institute for the following support: HST GO-07505.02A; HST GO-08177.06
(the High-Z Supernova Team survey); HST GO-08641.07A was provided by
NASA through a grant from the Space Telescope Science Institute, which
is operated by the Association of Universities for Research in
Astronomy, Inc., under NASA contract NAS5-26555.  We thank STScI for
the salary support for P.C. from grants GO-09114 and GO-09421. We
thank Stefano Benetti and Peter Meikle for many useful discussions
about SN~2002bo and for sharing their optical and IR
spectra. K.K. thanks LCO and NOAO for funding part of his postdoctoral
position.  Support for this work was provided to M.H. by NASA through
Hubble Fellowship grant HST HF-01139.01A, awarded by the Space Telescope
Science Institute.  J. L. P. and N. B. S. thank Armin Rest for 
the idea of a continuous \dmm\ parameterization.  We thank Lisa
Germany for making the mosaics of the LCO 2.5-m images of SN~1999ek.

We thank Michael Merrill and Richard Joyce for useful discussions
relating to the reduction of the SN 1991bg data obtained with
SQIID; and Gajus Miknaitis for help with reading two rather old
data tapes.

We are grateful for access to telescopes at
\facility{Cerro Tololo Inter-American Observatory},
\facility{Kitt Peak National Observatory},
\facility{Las Campanas Observatory},
\facility{Steward Observatory}, and the
\facility{Fred L. Whipple Observatory}.

\clearpage

\begin{deluxetable}{cccccccc}
\label{tab01}
\tablecolumns{8}
\rotate
\tablewidth{0pc}
\tablecaption{Optical Photometry of Supernova Field Stars}
\startdata
SN field   & star  &  $U$   &   $B$   &  $V$  &  $R$  & $I$  & N\tablenotemark{a} \\ \tableline \tableline
1999ek & 1 & \nodata &  14.766 (0.007) & 13.909 (0.011) & 13.361 (0.006) & 12.807 (0.006) & 3 \\
\ldots & 2 & \nodata &  15.863 (0.009) & 15.017 (0.009) & 14.463 (0.006) & 13.902 (0.009) & 3 \\
\ldots & 3 & \nodata &  17.117 (0.013) & 16.276 (0.011) & 15.723 (0.007) & 15.168 (0.009) & 3 \\
\ldots & 4 & \nodata &  14.266 (0.007) & 13.460 (0.009) & 12.935 (0.006) & 12.408 (0.008) & 3 \\
\ldots & 5 & \nodata &  13.592 (0.006) & 12.855 (0.010) & 12.374 (0.009) & 11.878 (0.011) & 3 \\
\ldots & 6 & \nodata &  16.106 (0.006) & 14.600 (0.010) & 13.761 (0.006) & 12.979 (0.006) & 3 \\
\ldots & 7 & \nodata &  14.420 (0.007) & 13.661 (0.010) & 13.169 (0.006) & 12.674 (0.007) & 3 \\
\ldots & 8 & \nodata &  16.249 (0.006) & 15.325 (0.008) & 14.729 (0.006) & 14.128 (0.006) & 3 \\
\ldots & 9 & \nodata &  15.492 (0.006) & 14.168 (0.011) & 13.387 (0.006) & 12.635 (0.008) & 3 \\
\ldots & 10 & \nodata & 15.273 (0.008) & 14.196 (0.009) & 13.525 (0.006) & 12.877 (0.006) & 3 \\
\\
2001bt & 1 &  15.964 (0.068) & 15.326 (0.006) & 14.424 (0.005) & 13.918 (0.012) & 13.509 (0.005) & 2 \\      
\ldots   & 2 &  16.840 (0.076) & 15.988 (0.011) & 14.938 (0.009) & 14.339 (0.012) & 13.828 (0.003) & 2 \\
\ldots   & 3 &  15.937 (0.068) & 15.852 (0.007) & 15.257 (0.007) & 14.881 (0.013) & 14.553 (0.007) & 2 \\
\ldots   & 4 &   \nodata       & 17.155 (0.007) & 16.288 (0.017) & 15.799 (0.016) & 15.402 (0.007) & 2 \\
\ldots   & 5 &  15.870 (0.068) & 15.427 (0.004) & 14.622 (0.005) & 14.165 (0.011) & 13.778 (0.007) & 2 \\
\ldots   & 6 &  15.505 (0.066) & 14.463 (0.004) & 13.333 (0.003) & 12.710 (0.011) & 12.167 (0.006) & 2 \\
\\
2001cn & 2 & 15.372 (0.011) & 15.004 (0.006) & 14.393 (0.004) & 13.674 (0.007) & 14.284	(0.059)	& 4 \\
\ldots & 3 & 16.145 (0.009) & 15.698 (0.007) & 14.891 (0.004) & 14.037 (0.007) & 14.445	(0.063)	& 4 \\
\ldots & 4 & 16.556 (0.011) & 16.174 (0.008) & 15.522 (0.004) & 14.769 (0.008) & 15.418	(0.064)	& 4 \\
\ldots & 5 & 16.577 (0.013) & 15.982 (0.011) & 14.925 (0.006) & 13.792 (0.012) & 13.944	(0.068)	& 4 \\
\ldots & 6 & 16.415 (0.011) & 16.020 (0.009) & 15.399 (0.006) & 14.639 (0.011) & 15.157	(0.061)	& 4 \\
\ldots & 7 & 17.378 (0.011) & 16.629 (0.010) & 15.457 (0.005) & 14.056 (0.010) &  \nodata     	& 4 \\
\ldots & 8 & 17.495 (0.015) & 17.083 (0.011) & 16.401 (0.005) & 15.611 (0.010) &  \nodata       & 4 \\
\\
2001cz & 1 & 15.490 (0.015) & 15.031 (0.003) & 14.173 (0.003) & 13.696 (0.004) & 13.265 (0.004)	& 7 \\
\ldots & 2 & 14.907 (0.014) & 14.716 (0.003) & 13.990 (0.004  & 13.568 (0.005) & 13.166 (0.003)	& 7 \\
\ldots & 3 & 17.004 (0.018) & 16.784 (0.006) & 16.027 (0.003) & 15.575 (0.003) & 15.153 (0.003)	& 7 \\
\ldots & 4 & 17.306 (0.021) & 16.946 (0.007) & 16.136 (0.004) & 15.672 (0.004) & 15.245 (0.004)	& 7 \\
\ldots & 5 & 16.985 (0.017) & 16.965 (0.007) & 16.278 (0.004) & 15.848 (0.003) & 15.431 (0.004)	& 7 \\
\ldots & 6 & 15.970 (0.018) & 15.882 (0.004) & 15.195 (0.003) & 14.784 (0.004) & 14.394 (0.003)	& 7 \\
\\
2002bo & 1 & 15.715 (0.034) & 15.252 (0.016) & 14.435 (0.012) & 14.000 (0.015) & 13.597 (0.014) &  7 \\
\ldots & 2 & 19.643 (0.119) & 18.529 (0.021) & 17.213 (0.019) & 16.391 (0.020) & 15.666 (0.023) &  5 \\
\ldots & 3 & 15.938 (0.028) & 15.691 (0.016) & 14.929 (0.013) & 14.492 (0.018) & 14.076 (0.015) &  7 \\
\ldots & 4 & 13.250 (0.019) & 13.046 (0.017) & 12.414 (0.015) & 12.066 (0.015) & 11.703 (0.015) &  3 \\
\ldots & 5 & 15.224 (0.038) & 15.072 (0.024) & 14.366 (0.020) & 13.970 (0.025) & 13.613 (0.021) &  3 \\
\ldots & 6 & 16.253 (0.041) & 16.291 (0.021) & 15.644 (0.019) & 15.278 (0.023) & 14.922 (0.021) &  4 \\
\enddata
\tablenotetext{a}{The number of nights of calibration measurements for a given star.
Values in parentheses in this table (and others) are 1-$\sigma$ error bars.}
\end{deluxetable}

\begin{deluxetable}{cccccc}
\label{tab02}
\tablecolumns{6}
\tablewidth{0pc} 
\tablecaption{Infrared Photometry of Supernova Field Stars}
\startdata 
Field   &    star  &   $J/J_s$   &  $H$  &  $K/K_s$\tablenotemark{a}  & N$_{obs}$\tablenotemark{b} \\ 
\tableline \tableline
1991bg  &    A     & 11.803 (0.010) & 11.164 (0.008) & 11.002 (0.028) & 1 1 1 \\
\ldots  &    J     & 10.822 (0.005) & 10.585 (0.005) & 10.574 (0.011) & 1 1 1 \\

1999ek  &    2     & 13.128 (0.006) & 12.738 (0.008) & 12.649 (0.022)   & 5 6 0 \\
\ldots  &    3    & 14.371 (0.015) & 13.978 (0.005) & 13.805 (0.044)   & 5 6 0 \\
\ldots  &    IR1  & 14.049 (0.012) & 13.302 (0.004) & 13.100 (0.028)  & 5 6 0 \\
2001bt  &    1    & 12.878 (0.005) & 12.492 (0.010) & 12.380 (0.009) & 2 2 2 \\
2001cn  &    IR1  & 13.771 (0.013) & 13.100 (0.011) & \nodata        & 1 1 0 \\
\ldots  &    3    & 13.478 (0.010) & 13.108 (0.009) & 13.046 (0.029) & 1 1 1 \\
\ldots  &    4    & 14.328 (0.012) & 14.011 (0.012) & 13.929 (0.037) & 1 1 1 \\
\ldots  &    7    & 13.196 (0.010) & 12.542 (0.009) & 12.394 (0.027) & 1 1 1 \\
\ldots  &    8    & 15.170 (0.016) & 14.781 (0.017) & \nodata        & 1 1 0 \\
2001cz  &    6    & 13.846 (0.055) & 13.547 (0.011) & 13.481 (0.027) & 2 1 1 \\
2002bo  &    1    & 13.051 (0.009) & 12.688 (0.008) & 12.676 (0.035) & 4 4 4 \\
\enddata
\tablenotetext{a} {The K-band values for the 1999ek stars are from 2MASS.}
\tablenotetext{b} {Number of nights of observations that were used to determine the
weighted means, for each of the three filters.}
\end{deluxetable}


\begin{deluxetable}{llccccc}
\label{tab03}
\tablecolumns{7}
\tablewidth{0pc} 
\tablecaption{Infrared Photometry of Type Ia Supernova Prototypes}
\startdata
Object & UT Date &   JD\tablenotemark{a}  & J & H & K & Notes\tablenotemark{b}\\
\tableline \tableline
SN 1991T  & 1991 Apr 17 &   364.46  & 11.73        &  11.16        & 10.93        & 1 \\
\ldots    & 1991 May 4  &   380.75  & 12.45 (0.09) &  12.44 (0.10) & 12.28 (0.10) & 2 \\
\ldots    & 1991 May 5  &   381.80  & 12.58 (0.10) &  12.29 (0.10) & 12.19 (0.10) & 2 \\
\ldots    & 1991 Jun 22 &   429.86  & 14.97 (0.30) &  13.72 (0.14) & 13.71 (0.25) & 2,3 \\
SN 1991bg & 1991 Dec 15 &   606.05  & 13.49 (0.03) &  13.45 (0.04) & 13.49 (0.05) & 4 \\
\ldots    & 1991 Dec 16 &   606.06  & 13.51 (0.03) &  13.44 (0.03) & 13.41 (0.05) & 4 \\
\ldots    & 1991 Dec 17 &   607.05  & 13.59 (0.04) & \nodata & \nodata & 4 \\
\enddata
\tablenotetext{a} {Julian Date {\em minus} 2,448,000.}
\tablenotetext{b} {1 = \citet{Men_Car91}; may include a significant amount of
host galaxy light. 2 = data on MSSSO IR system \citep{McG94}.  3 = \citet{Har_Str91}.
4 = data on Caltech IR system \citep{Eli_etal82}.}
\end{deluxetable}

\begin{deluxetable}{ccccc}
\label{tab04}
\tablecolumns{5}
\tablewidth{0pc}
\tablecaption{Optical Photometry of SN 1999ek}
\startdata
JD\tablenotemark{a} &  $B$   &  $V$  &  $R$  & $I$   \\ \tableline \tableline
478.840 & 18.061 (0.023) &  17.401 (0.018) &  16.901 (0.015) &  16.557 (0.016) \\
478.840 & 18.048 (0.021) &  17.378 (0.016) &  16.883 (0.015) &  16.569 (0.015) \\
479.836 & 17.987 (0.041) &  17.362 (0.035) &  16.778 (0.023) &  16.479 (0.027) \\
479.840 & 17.925 (0.037) &  17.351 (0.036) &  16.770 (0.024) &  16.520 (0.032) \\
480.758 & 18.070 (0.070) &  17.324 (0.058) &  16.760 (0.055) &  16.574 (0.079) \\
480.758 & 17.984 (0.082) &  17.339 (0.060) &  16.856 (0.058) &  16.573 (0.072) \\
481.785 & 17.938 (0.016) &  17.250 (0.015) &  16.807 (0.015) &  16.574 (0.015) \\
481.789 & 17.930 (0.023) &  17.278 (0.015) &  16.820 (0.015) &  16.580 (0.015) \\
483.863 & 17.955 (0.015) &  17.250 (0.015) &  16.817 (0.015) &  16.675 (0.016) \\
483.867 & 17.913 (0.014) &  17.240 (0.015) &  16.824 (0.015) &  16.681 (0.015) \\
\\
484.734 & 17.968 (0.014) &  17.243 (0.015) &  16.834 (0.015) &  16.705 (0.015) \\
484.738 & 17.983 (0.014) &  17.268 (0.015) &  16.850 (0.015) &  16.699 (0.015) \\
485.801 & 17.998 (0.029) &  17.300 (0.043) &     \nodata     &  16.751 (0.016) \\
485.805 & 18.034 (0.057) &  17.291 (0.023) &  16.816 (0.015) &  16.795 (0.023) \\
486.824 & 18.069 (0.014) &  17.275 (0.015) &  16.875 (0.015) &  16.788 (0.015) \\
486.824 & 18.062 (0.015) &  17.284 (0.015) &  16.870 (0.015) &  16.765 (0.015) \\
487.812 & 18.108 (0.014) &  17.325 (0.015) &  16.902 (0.015) &  16.851 (0.016) \\
487.812 & 18.103 (0.016) &  17.317 (0.015) &  16.901 (0.015) &  16.814 (0.017) \\
488.813 & 18.182 (0.014) &  17.364 (0.015) &  16.978 (0.015) &  16.917 (0.015) \\
489.762 & 18.289 (0.019) &  17.394 (0.021) &  17.032 (0.020) &  16.961 (0.024) \\
\\
489.766 & 18.258 (0.018) &  17.410 (0.019) &  17.044 (0.016) &  16.966 (0.024) \\
490.844 & 18.358 (0.014) &  17.455 (0.015) &  17.128 (0.015) &  17.096 (0.023) \\
490.848 & 18.374 (0.017) &  17.468 (0.015) &  17.138 (0.017) &  17.058 (0.019) \\
492.859 & 18.552 (0.018) &  17.583 (0.015) &  17.301 (0.015) &  17.223 (0.018) \\
492.871 & 18.541 (0.051) &  17.562 (0.024) &  17.298 (0.029) &  17.222 (0.023) \\
493.812 & 18.668 (0.016) &  17.647 (0.015) &  17.354 (0.015) &  17.287 (0.018) \\
493.828 & 18.672 (0.017) &  17.654 (0.015) &  17.350 (0.015) &  17.230 (0.018) \\
501.848 & 19.654 (0.026) &  18.116 (0.016) &  17.547 (0.016) &  17.175 (0.016) \\
501.852 & 19.671 (0.031) &  18.131 (0.018) &  17.526 (0.015) &  17.162 (0.017) \\
503.695 & 20.384 (0.208) &  18.300 (0.022) &  17.663 (0.016) &  \nodata   \\
\\
504.707 & 20.057 (0.176) &  18.248 (0.025) &  17.614 (0.017) &  17.065 (0.016) \\
506.778 & 20.308 (0.206) &  \nodata        &    \nodata      &  \nodata   \\
508.836 & 20.312 (0.134) &  18.518 (0.027) &  17.722 (0.018) &  17.075 (0.015) \\
\enddata
\tablenotetext{a} {Julian Date $minus$ 2,451,000.}
\end{deluxetable}

\begin{deluxetable}{ccccc}
\label{tab05}
\tablecolumns{5}
\tablewidth{0pc}
\tablecaption{Infrared Photometry of SN 1999ek}
\startdata
JD\tablenotemark{a}  &   $J$ &  $H$  &  $K$ & Telescope\tablenotemark{b}  \\ \tableline \tableline
473.99 &  \nodata       & 16.414 (0.115) & 16.544 (0.076) & 1 \\
477.67 & 16.150 (0.039) & 16.079 (0.040) & \nodata & 2 \\
477.80 & 16.124 (0.020) & 16.281 (0.018) & \nodata & 3 \\
478.69 & 16.097 (0.050) & 16.136 (0.050) & \nodata & 2 \\
481.82 & 16.209 (0.019) & 16.331 (0.020) & \nodata & 3 \\
483.78 & 16.415 (0.023) & 16.464 (0.038) & \nodata & 4 \\
484.81 & 16.436 (0.024) & 16.485 (0.024) & \nodata & 4 \\
485.78 & 16.539 (0.024) & 16.405 (0.163) & \nodata & 4 \\
486.82 & 16.908 (0.068) & 16.676 (0.135) & \nodata & 4 \\
487.79 & 16.778 (0.023) & 16.648 (0.025) & \nodata & 4 \\
488.80 & 16.935 (0.022) & 16.645 (0.029) & \nodata & 4 \\
489.84 & 17.134 (0.023) & 16.525 (0.031) & \nodata & 4 \\
497.66 & 17.819 (0.081) & 16.455 (0.070) & \nodata & 2 \\
503.64 & 17.563 (0.065) & 16.296 (0.059) & \nodata & 2 \\
504.67 & 17.441 (0.088) & 16.253 (0.073) & \nodata & 2 \\
\enddata
\tablenotetext{a} {Julian Date $minus$ 2,451,000.}
\tablenotetext{b} {1 = FLWO 1.2-m; 2 = Steward Observatory 90-inch;
3 = LCO 2.5-m; 4 = LCO 1-m.}
\end{deluxetable}

\begin{deluxetable}{cccccc}
\label{tab06}
\tablecolumns{6}
\tablewidth{0pc}
\tablecaption{Optical Photometry of SN 2001bt}
\startdata
JD\tablenotemark{a} & $B$   &  $V$  &  $R$  & $I$ & Telescope\tablenotemark{b}  \\ \tableline \tableline
2055.78 &  16.260 (0.023) &  16.096 (0.009) &  15.817 (0.015) &   \nodata       & 2 \\
2061.90 &  15.577 (0.014) &  15.414 (0.006) &  15.212 (0.010) &  15.305 (0.015) & 2 \\
2061.82 &  15.571 (0.011) &  15.414 (0.007) &  15.210 (0.016) &  15.289 (0.013) & 1 \\
2064.77 &  15.522 (0.039) &  15.319 (0.019) &  15.147 (0.073) &  15.359 (0.042) & 1 \\
2067.83 &  15.668 (0.012) &  15.317 (0.008) &  15.155 (0.018) &  15.460 (0.014) & 1 \\
2070.82 &  15.856 (0.019) &  15.393 (0.011) &  15.266 (0.026) &  15.567 (0.019) & 1 \\
2073.92 &  16.197 (0.030) &  15.553 (0.018) &     \nodata     &  15.760 (0.035) & 1 \\
2076.86 &  16.503 (0.020) &  15.733 (0.008) &  15.663 (0.023) &  15.909 (0.025) & 1 \\
2079.89 &  16.945 (0.030) &  15.967 (0.017) &  15.778 (0.028) &  15.880 (0.035) & 1 \\
2082.75 &  17.306 (0.021) &  16.112 (0.011) &  15.809 (0.022) &  15.771 (0.019) & 1 \\
2085.74 &  17.689 (0.048) &  16.301 (0.026) &  15.869 (0.042) &  15.761 (0.039) & 1 \\
2088.70 &  17.937 (0.033) &  16.451 (0.019) &  16.006 (0.031) &  15.747 (0.031) & 1 \\
2091.68 &  18.156 (0.045) &  16.678 (0.017) &  16.144 (0.031) &  15.794 (0.026) & 1 \\
2094.81 &  18.410 (0.178) &  16.875 (0.047) &    \nodata      &  16.044 (0.079) & 1 \\
2096.78 &    \nodata      &  16.978 (0.035) &  16.529 (0.070) &  16.238 (0.063) & 1 \\
2097.67 &  18.529 (0.088) &  17.102 (0.028) &  16.566 (0.051) &  16.200 (0.043) & 1 \\
2101.75 &  18.574 (0.161) &  17.409 (0.061) &    \nodata      &  16.493 (0.105) & 1 \\
2106.64 &  18.791 (0.088) &  17.438 (0.041) &  17.005 (0.067) &  16.733 (0.061) & 1 \\
2115.86 &  18.904 (0.039) &  17.663 (0.014) &  17.285 (0.025) &  17.055 (0.030) & 1 \\
2122.67 &  18.923 (0.099) &  17.823 (0.027) &  17.453 (0.060) &  17.345 (0.084) & 1 \\
2123.65 &  18.920 (0.051) &  17.869 (0.011) &  17.507 (0.024) &  17.403 (0.027) & 1 \\
2128.75 &  19.121 (0.077) &  17.958 (0.041) &  17.626 (0.060) &  17.605 (0.062) & 1 \\
2131.76 &  19.014 (0.041) &  18.048 (0.013) &  17.750 (0.025) &  17.762 (0.042) & 1 \\
\enddata
\tablenotetext{a} {Julian Date $minus$ 2,450,000.}
\tablenotetext{b} {1 = YALO 1-m; 2 = CTIO 1.5-m.}
\end{deluxetable}

\begin{deluxetable}{ccccc}
\label{tab07}
\tablecolumns{5}
\tablewidth{0pc}
\tablecaption{Infrared Photometry of SN 2001bt}
\startdata
JD\tablenotemark{a} &  $J$ &  $H$  &  $K$  & Telescope\tablenotemark{b}  \\ \tableline \tableline
2061.81 &  15.472 (0.029) &  15.695 (0.061) &  15.562 (0.078) & 1 \\
2061.85 &  15.420 (0.015) &  15.645 (0.020) &  15.399 (0.026) & 2 \\ 
2064.76 &  15.342 (0.073) &  15.895 (0.160) &  15.629 (0.275) & 1 \\ 
2068.83 &  16.163 (0.041) &  16.120 (0.051) &    \nodata      & 2 \\ 
2069.83 &  16.291 (0.036) &  16.163 (0.045) &  15.653 (0.030) & 2 \\  
2070.81 &  16.244 (0.051) &  16.087 (0.091) &  15.944 (0.208) & 1 \\ 
2073.92 &  17.029 (0.118) &  16.226 (0.190) &  15.860 (0.217) & 1 \\
2076.86 &  17.168 (0.071) &  16.221 (0.114) &  15.875 (0.126) & 1 \\ 
2079.89 &  17.010 (0.088) &  16.048 (0.143) &  15.776 (0.151) & 1 \\
2082.75 &  17.051 (0.062) &  15.930 (0.094) &    \nodata      & 1 \\
2088.70 &  16.649 (0.042) &  15.669 (0.066) &  15.743 (0.096) & 1 \\
2091.67 &  16.435 (0.037) &  15.783 (0.065) &  15.827 (0.102) & 1 \\
2094.80 &  16.633 (0.089) &  15.934 (0.199) &    \nodata      & 1 \\
2096.78 &  16.733 (0.081) &  16.077 (0.130) &    \nodata      & 1 \\
2097.66 &  16.785 (0.047) &  16.156 (0.079) &    \nodata      & 1 \\
2101.70 &  17.355 (0.163) &  16.254 (0.251) &  17.003 (0.538) & 1 \\
2106.63 &  17.677 (0.078) &  16.578 (0.123) &  17.276 (0.259) & 1 \\
2115.84 &  18.052 (0.116) &  16.942 (0.184) &   \nodata       & 1 \\
2122.65 &  18.904 (0.234) &  17.227 (0.342) &   \nodata       & 1 \\
2123.63 &  18.747 (0.149) &  17.275 (0.221) &   \nodata       & 1 \\
2128.73 &  19.428 (0.277) &  17.450 (0.401) &   \nodata       & 1 \\
\enddata
\tablenotetext{a} {Julian Date $minus$ 2,450,000.}
\tablenotetext{b} {1 = YALO 1-m; 2 = LCO 2.5-m.}
\end{deluxetable}

\begin{deluxetable}{ccccccc}
\label{tab08}
\tablecolumns{7}
\rotate
\tablewidth{0pc}
\tablecaption{Optical Photometry of SN 2001cn}
\startdata
JD\tablenotemark{a} &  $U$ & $B$   &  $V$  &  $R$  & $I$ & Telescope\tablenotemark{b}  \\ 
\tableline \tableline
2075.85 &  \nodata       &  15.559 (0.013) &  15.312 (0.008) &  15.162 (0.011) &  15.470 (0.013) &  2 \\
2075.84 &  \nodata       &  15.639 (0.014) &  15.289 (0.007) &  15.182 (0.016) &  15.479 (0.016) &  1 \\
2076.80 & 15.716 (0.057) &  15.611 (0.022) &  15.332 (0.013) &  15.196 (0.020) &  15.524 (0.022) &  2 \\
2076.85 &  \nodata       &  15.631 (0.023) &  15.312 (0.015) &  15.204 (0.024) &  15.582 (0.024) &  1 \\
2077.79 & 15.805 (0.059) &  15.676 (0.023) &  15.365 (0.013) &  15.233 (0.020) &  15.573 (0.023) &  2 \\
2078.80 & 15.868 (0.062) &  15.759 (0.022) &  15.410 (0.013) &  15.300 (0.019) &  15.650 (0.023) &  2 \\
2079.73 &  \nodata       &  15.785 (0.015) &  15.411 (0.010) &  15.350 (0.019) &  15.757 (0.018) &  1 \\
2082.71 &  \nodata       &  16.099 (0.008) &  15.556 (0.005) &  15.556 (0.008) &  15.989 (0.010) &  1 \\
2085.71 &  \nodata       &  16.425 (0.009) &  15.765 (0.005) &  15.739 (0.010) &  16.084 (0.033) &  1 \\
2088.67 &  \nodata       &  16.822 (0.011) &  15.930 (0.006) &  15.817 (0.010) &  16.041 (0.014) &  1 \\
\\
2091.65 &  \nodata       &  17.173 (0.019) &  16.103 (0.005) &  15.865 (0.008) &  15.941 (0.010) &  1 \\
2094.81 &  \nodata       &  17.532 (0.020) &  16.264 (0.005) &  15.929 (0.010) &  15.874 (0.015) &  1 \\
2096.72 &  \nodata       &  17.729 (0.059) &  16.394 (0.033) &  15.972 (0.049) &  15.809 (0.061) &  1 \\ 
2097.71 &  \nodata       &  17.800 (0.021) &  16.435 (0.009) &  16.007 (0.015) &  15.836 (0.029) &  1 \\
2101.72 &  \nodata       &  18.086 (0.018) &  16.724 (0.010) &  16.232 (0.017) &  15.903 (0.017) &  1 \\
2104.67 &  \nodata       &  18.275 (0.018) &  16.901 (0.006) &  16.435 (0.011) &  16.085 (0.011) &  1 \\
2106.67 &  \nodata       &  18.379 (0.016) &  17.029 (0.006) &  16.569 (0.011) &  16.262 (0.017) &  1 \\
2114.86 &  \nodata       &  18.667 (0.026) &  17.378 (0.010) &  16.995 (0.025) &  16.787 (0.027) &  1 \\
2117.81 &  \nodata       &  18.664 (0.033) &  17.461 (0.012) &  17.071 (0.023) &  16.903 (0.081) &  1 \\
2122.62 &  \nodata       &  18.763 (0.071) &  17.634 (0.030) &  17.201 (0.047) &  17.117 (0.053) &  1 \\
\\
2123.61 &  \nodata       &  18.757 (0.035) &  17.613 (0.011) &  17.255 (0.019) &  17.141 (0.020) &  1 \\
2127.67 &  \nodata       &  18.846 (0.075) &  17.747 (0.038) &  17.399 (0.057) &  17.309 (0.061) &  1 \\
2130.76 &  \nodata       &  18.897 (0.056) &  17.763 (0.030) &  17.463 (0.044) &  17.408 (0.066) &  1 \\
2140.72 &  \nodata       &  19.057 (0.054) &  18.059 (0.022) &  17.794 (0.035) &  17.774 (0.046) &  1 \\
2156.71 &  \nodata       &  19.127 (0.080) &  18.546 (0.040) &  18.278 (0.077) &  18.361 (0.111) &  1 \\
\enddata
\tablenotetext{a} {Julian Date $minus$ 2,450,000.}
\tablenotetext{b} {1 = YALO 1-m; 2 = CTIO 0.9-m.}
\end{deluxetable}

\begin{deluxetable}{ccccc}
\label{tab09}
\tablecolumns{4}
\tablewidth{0pc}
\tablecaption{YALO Infrared Photometry of SN 2001cn}
\startdata
JD\tablenotemark{a} &  $J$ &  $H$  &  $K$   \\ \tableline \tableline
2075.84 &  15.899 (0.028) &  16.084 (0.062) &  15.667 (0.095) \\
2076.84 &  16.171 (0.046) &  16.144 (0.071) &  15.684 (0.099) \\
2079.73 &  16.561 (0.044) &  16.088 (0.077) &  15.829 (0.104) \\
2082.71 &  17.011 (0.035) &  16.164 (0.058) & \nodata \\
2085.69 &  17.379 (0.059) &  16.039 (0.095) & \nodata \\
2088.67 &  17.191 (0.036) &  15.907 (0.056) & \nodata \\
2091.64 &  17.130 (0.036) &  15.829 (0.057) & \nodata \\
2094.74 &  16.959 (0.036) &  15.873 (0.071) & \nodata \\
2096.72 &  16.842 (0.040) &  15.707 (0.069) & \nodata \\
2097.69 &  16.831 (0.034) &  15.665 (0.054) & \nodata \\
2101.70 &  16.560 (0.031) &  15.792 (0.052) & \nodata \\
2104.66 &  16.629 (0.029) &  15.895 (0.048) & \nodata \\
2106.66 &  16.796 (0.030) &  16.072 (0.049) & \nodata \\
2114.84 &  17.716 (0.085) &  16.606 (0.136) & \nodata \\
2117.47 &  17.896 (0.079) &  16.808 (0.128) & \nodata \\
2122.60 &  18.273 (0.153) &  16.887 (0.232) & \nodata \\
2123.60 &  18.400 (0.103) &  16.905 (0.156) & \nodata \\
2127.66 &  18.401 (0.135) &  17.445 (0.216) & \nodata \\
2130.76 &  18.855 (0.193) &  17.449 (0.301) & \nodata \\
\enddata
\tablenotetext{a} {Julian Date $minus$ 2,450,000.}
\end{deluxetable}

\begin{deluxetable}{ccccccc}
\label{tab10}
\tablecolumns{7}
\rotate
\tablewidth{0pc}
\tablecaption{Optical Photometry of SN 2001cz}
\startdata
JD\tablenotemark{a} &  $U$ & $B$   &  $V$  &  $R$  & $I$   & Telescope\tablenotemark{b}\\ 
\tableline \tableline
2097.50	&  15.297 (0.045)& 15.740 (0.037) & 15.682 (0.034) & 15.489 (0.040) & 15.474 (0.037) & 2 \\
2101.53	&  \nodata       & 15.465 (0.012) & 15.307 (0.006) & 15.202 (0.015) & 15.304 (0.013) & 1 \\
2104.47	&  \nodata       & 15.436 (0.016) & 15.240 (0.010) & 15.136 (0.018) & 15.378 (0.017) & 1 \\
2106.47	&  \nodata       & 15.474 (0.010) & 15.264 (0.006) & 15.160 (0.014) & 15.432 (0.013) & 1 \\
2112.53	&  \nodata       & 15.843 (0.022) & 15.449 (0.013) & 15.450 (0.028) & 15.767 (0.028) & 1 \\
2114.48	&  \nodata       & 15.957 (0.018) & 15.535 (0.012) & 15.555 (0.023) & 15.885 (0.022) & 1 \\
2117.48	&  \nodata       & 16.237 (0.014) & 15.735 (0.006) & 15.721 (0.018) & 16.081 (0.017) & 1 \\
2123.47	&  \nodata       & 16.944 (0.021) & 16.061 (0.009) & 15.855 (0.022) & 15.973 (0.025) & 1 \\
2128.48	&  \nodata       & 17.488 (0.070) & 16.291 (0.019) & 15.931 (0.033) & 15.872 (0.032) & 1 \\
2130.49	&  \nodata       & 17.694 (0.040) & 16.432 (0.024) & 16.008 (0.039) & 15.826 (0.036) & 1 \\
2137.49	&  \nodata       & 18.203 (0.085) & 16.858 (0.012) & \nodata        & \nodata	     & 1 \\
2144.48	&  \nodata       & 18.462 (0.041) & 17.195 (0.008) & 16.767 (0.033) & 16.462 (0.025) & 1 \\
2151.49	&  \nodata       & 18.608 (0.040) & 17.457 (0.018) & 17.136 (0.031) & 16.801 (0.033) & 1 \\
2155.51	&  \nodata       & 18.623 (0.164) & 17.591 (0.045) & \nodata        & \nodata	     & 1 \\
\enddata
\tablenotetext{a} {Julian Date $minus$ 2,450,000.}
\tablenotetext{b} {1 = YALO 1.0-m; 2 = CTIO 0.9-m}
\end{deluxetable}

\begin{deluxetable}{ccccc}
\label{tab11}
\tablecolumns{4}
\tablewidth{0pc}
\tablecaption{YALO Infrared Photometry of SN 2001cz}
\startdata
JD\tablenotemark{a} &  $J$ &  $H$  &  $K$    \\ \tableline \tableline
2101.52 &  15.404 (0.061) &  15.730 (0.126) &  15.397 (0.132) \\
2104.46 &  15.499 (0.061) &  15.879 (0.121) &  15.607 (0.130) \\
2106.46 &  15.623 (0.064) &  15.996 (0.121) &  15.685 (0.148) \\
2112.52 &  16.580 (0.120) &  16.023 (0.216) &  15.759 (0.258) \\
2114.47 &  16.806 (0.085) &  16.153 (0.155) &  15.805 (0.172) \\
2117.46 &  17.163 (0.099) &  16.155 (0.168) &  16.062 (0.187) \\
2123.46 &  17.477 (0.125) &  15.882 (0.198) &  15.785 (0.208) \\
2128.47 &  17.029 (0.091) &  15.646 (0.155) &  15.642 (0.177) \\
2130.48 &  16.833 (0.079) &  15.646 (0.140) &  15.704 (0.161) \\
2137.49 &  16.540 (0.102) &  15.722 (0.188) &  \nodata \\
2144.47 &  17.020 (0.078) &  16.220 (0.143) &  \nodata \\
2151.47 &  17.664 (0.098) &  16.574 (0.173) &  \nodata \\
\enddata
\tablenotetext{a} {Julian Date $minus$ 2,450,000.}
\end{deluxetable}

\begin{deluxetable}{cccccccccccc}
\label{tab12}
\tablecolumns{12}
\rotate
\tablewidth{0pc}
\tabletypesize{\scriptsize}
\tablecaption{Optical Photometry of SN 2002bo\tablenotemark{a}}
\startdata
JD\tablenotemark{b} & $U$ & $\Delta U$ & $B$ & $\Delta B$ & $V$ & $\Delta V$ & $R$ & $\Delta R$ & $I$ & $\Delta I$ & Telescope\tablenotemark{c} \\ 
\tableline \tableline
345.67 & 16.248 (0.094) &   [0.047] & 15.797 (0.016) &  $-$0.073 & 15.264 (0.013) & \phs0.015 & 14.910 (0.017) & \phs0.000 & 14.876 (0.026) & \phs0.030 & 1 \\
346.68 & 15.759 (0.086) &   [0.047] & 15.431 (0.009) &  $-$0.069 & 14.940 (0.011) & \phs0.014 & 14.582 (0.011) & \phs0.000 & 14.536 (0.023) & \phs0.023 & 1 \\
350.56 & 14.381 (0.064) &   [0.073] & 14.397 (0.020) &  $-$0.030 & 14.069 (0.010) & \phs0.003 & 13.812 (0.011) & \phs0.005 & 13.805 (0.013) &  $-$0.003 & 2 \\
350.66 & 14.655 (0.071) &   [0.047] & 14.484 (0.022) &  $-$0.040 & 14.099 (0.010) & \phs0.017 & 13.840 (0.016) & $-$0.010  & 13.784 (0.011) & \phs0.001 & 1 \\
354.65 & 14.313 (0.081) & $-$0.095  & 14.136 (0.014) &  $-$0.038 & 13.706 (0.009) & \phs0.027 & 13.556 (0.013) & $-$0.015  & 13.621 (0.010) &  $-$0.029 & 1 \\
358.64 & 14.448 (0.099) & $-$0.135  & 14.126 (0.010) &  $-$0.041 & 13.577 (0.013) & \phs0.028 & 13.501 (0.015) & $-$0.014  & 13.699 (0.019) &  $-$0.064 & 1 \\
360.65 & 14.351 (0.074) & \phs0.170 & 14.127 (0.047) &  $-$0.007 & 13.638 (0.040) & \phs0.000 & 13.508 (0.030) & \phs0.005 & 13.749 (0.030) &  $-$0.005 & 3 \\
361.66 & 14.412 (0.074) & \phs0.170 & 14.205 (0.040) &  $-$0.004 & 13.659 (0.040) & $-$0.001  & 13.488 (0.030) & \phs0.005 & 13.784 (0.030) &  $-$0.006 & 3 \\
362.62 & 14.837 (0.073) & $-$0.107  & 14.308 (0.012) &  $-$0.031 & 13.654 (0.010) & \phs0.032 & 13.574 (0.012) & $-$0.015  & 13.864 (0.032) &  $-$0.102 & 1 \\
366.62 & 15.298 (0.069) & $-$0.118  & 14.658 (0.025) &  $-$0.026 & 13.815 (0.014) & \phs0.036 & 13.818 (0.026) & \phs0.009 & 14.144 (0.033) &  $-$0.140 & 1 \\
370.55 & 15.838 (0.098) & $-$0.130  & 15.098 (0.013) &  $-$0.023 & 14.068 (0.014) & \phs0.040 & 14.047 (0.032) & \phs0.028 & 14.297 (0.034) &  $-$0.138 & 1 \\
373.60 & 15.810 (0.064) & \phs0.179 & 15.390 (0.020) & \phs0.021 & 14.349 (0.014) & $-$0.023  & 14.180 (0.011) & \phs0.008 & 14.349 (0.010) &  $-$0.003 & 2 \\
376.58 & 16.710 (0.114) & $-$0.147  & 15.758 (0.031) &  $-$0.020 & 14.413 (0.015) & \phs0.047 & 14.123 (0.029) & \phs0.039 & 14.098 (0.038) &  $-$0.111 & 1 \\
381.58 & 17.225 (0.157) & $-$0.162  & 16.293 (0.028) &  $-$0.072 & 14.658 (0.011) & \phs0.052 & 14.237 (0.022) & \phs0.024 & 14.045 (0.013) &  $-$0.088 & 1 \\
385.57 & 17.606 (0.249) & $-$0.174  & 16.587 (0.054) &  $-$0.080 & 14.896 (0.027) & \phs0.056 & 14.438 (0.026) & $-$0.001  & 14.066 (0.015) &  $-$0.070 & 1 \\
389.58 & 17.621 (0.215) & $-$0.185  & 16.858 (0.039) &  $-$0.078 & 15.218 (0.022) & \phs0.056 & 14.745 (0.014) & $-$0.009  & 14.346 (0.013) &  $-$0.060 & 1 \\
390.53 &   \nodata      & \nodata   & 16.875 (0.046) &  $-$0.076 & 15.270 (0.019) & \phs0.055 & 14.805 (0.014) & $-$0.008  & 14.429 (0.015) &  $-$0.059 & 1 \\
397.48 &   \nodata      & \nodata   & 17.112 (0.057) &  $-$0.067 & 15.548 (0.016) & \phs0.052 & 15.169 (0.015) & $-$0.005  & 14.874 (0.014) &  $-$0.047 & 1 \\
400.54 &   \nodata      & \nodata   & 17.147 (0.013) &  $-$0.062 & 15.660 (0.015) & \phs0.050 & 15.285 (0.018) & $-$0.001  & 14.976 (0.012) &  $-$0.041 & 1 \\
\enddata
\tablenotetext{a} {The corrections are to be {\em added} to the data.
This transforms the data to the filter system of \citet{Bes90}.
Values in square brackets could be in error by as much as 0.1 mag.}
\tablenotetext{b} {Julian Date $minus$ 2,452,000.}
\tablenotetext{c} {1 = YALO 1-m; 2 = CTIO 0.9-m; 3 = CTIO 1.5-m.}
\end{deluxetable}

\begin{deluxetable}{ccccccc}
\label{tab13}
\tablecolumns{7}
\tablewidth{0pc}
\tablecaption{YALO Infrared Photometry of SN 2002bo\tablenotemark{a}}
\startdata
JD\tablenotemark{b}  &   $J$ &  $\Delta J$ &  $H$  &  $\Delta H$ & $K$ & $\Delta K$  \\ \tableline \tableline
345.69 &  14.590 (0.030) &  \phs0.033 & 14.637 (0.039) & $-$0.024  & 14.765 (0.071) & \phs0.002 \\
346.68 &  14.442 (0.057) &  \phs0.036 & 14.493 (0.029) & $-$0.024  & 14.635 (0.059) & \phs0.002 \\
350.66 &  13.806 (0.084) &  \phs0.048 & 13.956 (0.028) & $-$0.026  & 13.977 (0.041) & \phs0.010 \\
354.65 &  13.619 (0.023) &  \phs0.061 & 13.907 (0.026) & $-$0.029  & 13.862 (0.041) & \phs0.021 \\
358.64 &  13.852 (0.030) &  \phs0.059 & 14.103 (0.030) & $-$0.032  & 13.984 (0.074) & \phs0.037 \\

362.62 &  14.262 (0.029) &  $-$0.005  & 14.251 (0.023) & $-$0.035  & 14.214 (0.051) & \phs0.055 \\
366.62 &  15.044 (0.099) &  $-$0.069  & 14.382 (0.054) & $-$0.038  & 14.482 (0.070) & \phs0.037 \\
370.55 &  15.472 (0.054) &  $-$0.086  & 14.276 (0.023) & $-$0.043  & 14.259 (0.061) & \phs0.024 \\
376.58 &  15.186 (0.075) &  $-$0.081  & 14.020 (0.032) & $-$0.064  & 14.090 (0.103) & \phs0.006 \\
381.58 &  14.828 (0.047) &  $-$0.076  & 13.890 (0.020) & $-$0.070  & 13.931 (0.040) & $-$0.011 \\

385.57 &  14.756 (0.043) &  $-$0.073  & 14.005 (0.029) & $-$0.052  & 14.101 (0.052) & $-$0.008 \\
389.57 &  14.905 (0.025) &  $-$0.086  & 14.122 (0.027) & $-$0.021  & 14.370 (0.048) & \phs0.001 \\
390.53 &  14.973 (0.038) &  $-$0.091  & 14.307 (0.027) & $-$0.015  & 14.530 (0.050) & \phs0.001 \\
397.48 &  15.731 (0.075) &  $-$0.124  & 14.663 (0.037) & \phs0.030 & 14.910 (0.071) & \phs0.002 \\
400.54 &  15.835 (0.037) &  $-$0.125  & 14.839 (0.030) & \phs0.033 & 15.158 (0.092) & \phs0.002 \\
\enddata
\tablenotetext{a} {To transform the data in columns 2, 4, and 6 to the
photometric system of \citet{Per_etal98} one {\em adds} the
corresponding corrections in columns 3, 5, and 7.}
\tablenotetext{b} {Julian Date $minus$ 2,452,000.}
\end{deluxetable}


\begin{deluxetable}{ccclrcl}
\label{tab14}
\tablecolumns{7}
\tablewidth{0pc} 
\tablecaption{Table of spectral observations}
\startdata 
Object & Telescope\tablenotemark{a} & Grating & UT Date &  Age\tablenotemark{b}
& Exptime\tablenotemark{c}  & Observer(s) \\ \tableline \tableline
SN~1999ee &  1 & 09 & 25.09 Oct 1999 &   +7.5  &  2400 &  C. Smith \\
  \ldots  &  1 & 13 & 25.18 Oct 1999 &   +7.6  &  1800 &  C. Smith \\
 \ldots   &  1 & 13 & 31.03 Oct 1999 &  +13.4  &  3600 &  R. Leiton, S. Pizarro \\
\hline
SN~1999ek &  1 & 13 & 25.37 Oct 1999 &  $-$5.2 &  1200 &  C. Smith \\
 \ldots   &  1 & 13 & 26.30 Oct 1999 &  $-$4.2 &  5400 &  P. Ugarte \\
 \ldots   &  1 & 13 & 30.33 Oct 1999 &  $-$0.2 &  3600 &  R. Leiton, S. Pizarro \\
 \ldots   &  1 & 13 & 01.32 Nov 1999 &    +1.8 &  3600 &  R. Leiton, S. Pizarro \\
\hline
SN~2001bt &  2 & grism & 26.37 May 2001 & $-$7.6 & 600 &   M. Phillips  \\
%
%
%
\hline
SN~2001cn &  1 & 09 & 14.30 Jun 2001 &      +3.4 & 1800 &  D. Norman, K. Olsen \\
  \ldots  &  1 & 09 & 22.31 Jun 2001 &     +11.2 &  600 &  J. Huchra \\
\enddata
\tablenotetext{a} {1 = CTIO 1.5-m; 2 = LCO 2.5-m.}
\tablenotetext{b} {Number of days in observer's frame since T($B_{max}$).}
\tablenotetext{c} {Total exposure time in seconds.}
\end{deluxetable}

\begin{deluxetable}{cccccccccc}
\label{tab15}
\tablecolumns{10}
\rotate
\tablewidth{0pc} 
\tablecaption{Useful Data for Supernovae}
\startdata 
SN    &  $v_{CMB}$\tablenotemark{a} & T($B_{max}$)\tablenotemark{b} &  
$B{max}$ & $V{max}$ & $R{max}$ & $I{max}$ & \dmm\  
& E($B-V$)$^{\tablenotemark{c}}_{Gal}$   & E($B-V$)$_{host}$  \\ \tableline \tableline
1999ek  & 5278  & 1482.04 (0.11)  & 17.91(0.01) & 17.25(0.01) & 16.88(0.01) & 16.56(0.01) & 1.17 (0.03)  & 0.561  & 0.203 (0.046) \\
2001bt  & 4332  & 2063.43 (0.22)  & 15.51(0.02) & 15.30(0.02) & 15.17(0.02) & 15.16(0.02) & 1.18 (0.03)  & 0.065  & 0.256 (0.018) \\
2001cn  & 4628  & 2071.56 (0.15)  & 15.40(0.01) & 15.27(0.01) & 15.20(0.01) & 15.26(0.01) & 1.15 (0.02)  & 0.059  & 0.165 (0.014) \\
2001cz  & 4900  & 2103.93 (0.12)  & 15.38(0.01) & 15.27(0.01) & 15.20(0.01) & 15.28(0.01) & 1.05 (0.03)  & 0.092  & 0.123 (0.017) \\
2002bo  & 1696  & 2356.50 (0.20)  & 14.02(0.02) & 13.47(0.01) & 13.61(0.01) & 13.55(0.02) & 1.12 (0.03)  & 0.025  & 0.365 (0.014) \\
\enddata
\tablenotetext{a} {Recession velocity in km s$^{-1}$ with respect to the Cosmic Microwave 
Background radiation.}
\tablenotetext{b} {Julian Date {\em minus} 2,450,000.}
\tablenotetext{c} {From \citet{Sch_etal98};  but see {\em caveat} from 
\citet{Arc_Goo99}.} 
\end{deluxetable}

\begin{deluxetable}{lclcccccccc}
\label{tab16}
\tablecolumns{11}
\rotate
\tablewidth{0pc}
\tablecaption{Infrared Apparent Magnitudes at Maximum and Extinction-Corrected Values\tablenotemark{a}}
\startdata
SN & A$_V$ & $J_{max}$ & $J_{corr}$ &
$H_{max}$ & $H_{corr}$ & $K_{max}$ & $K_{corr}$ & N$_{J,H,K}$\tablenotemark{b} \\
\hline \hline
1999ek  &  2.37(16) & 16.16(07) & 15.49(08) & 16.30(08) & 15.85(09) & 16.32(08) & 16.05(08) & 11,12,1 \\
2001bt  &  1.06(06) & 15.48(04) & 15.20(04) & 15.69(06) & 15.50(06) & 15.48(05) & 15.37(05) & 6,6,5 \\
2001cn  &  0.69(04) & 15.61(08) & 15.41(08) & 15.90(11) & 15.77(11) & 15.58(12) & 15.50(12) & 3,3,3 \\
2001cz  &  0.67(05) & 15.49(06) & 15.30(06) & 15.78(12) & 15.65(12) & 15.53(13) & 15.45(13) & 4,4,4 \\
2002bo  &  1.21(04) & 13.55(07) & 13.21(07) & 13.94(03) & 13.71(03) & 13.89(04) & 13.75(04) & 6,6,6 \\  
\enddata
\tablenotetext{a} {Values in parentheses are uncertainties in hundredths of a magnitude.}
\tablenotetext{b} {Number of data points prior to $t^{\prime}$ = +10 d in ``stretched time''.}
\end{deluxetable}

\begin{deluxetable}{ccccc}
\label{tab17}
\tablecolumns{4}
\tablewidth{0pc}
\tablecaption{Mean Absolute Magnitudes of Type Ia SNe at Maximum\tablenotemark{a}}
\startdata
Filter     &  $\langle$M$\rangle$ & $\sigma_x$ & $\chi^2_\nu$ & N     \\ \hline \hline
  J        &     $-$18.61(03)     &   $\pm$ 0.131  & 1.29 & 22  \\
  H        &     $-$18.28(03)     &   $\pm$ 0.148  & 1.52 & 21  \\
  K        &     $-$18.44(03)     &   $\pm$ 0.145  & 0.99 & 20  \\
\enddata
\tablenotetext{a} {For objects with \dmm\ $\lesssim$ 1.7.}
\end{deluxetable}

\clearpage

\figcaption[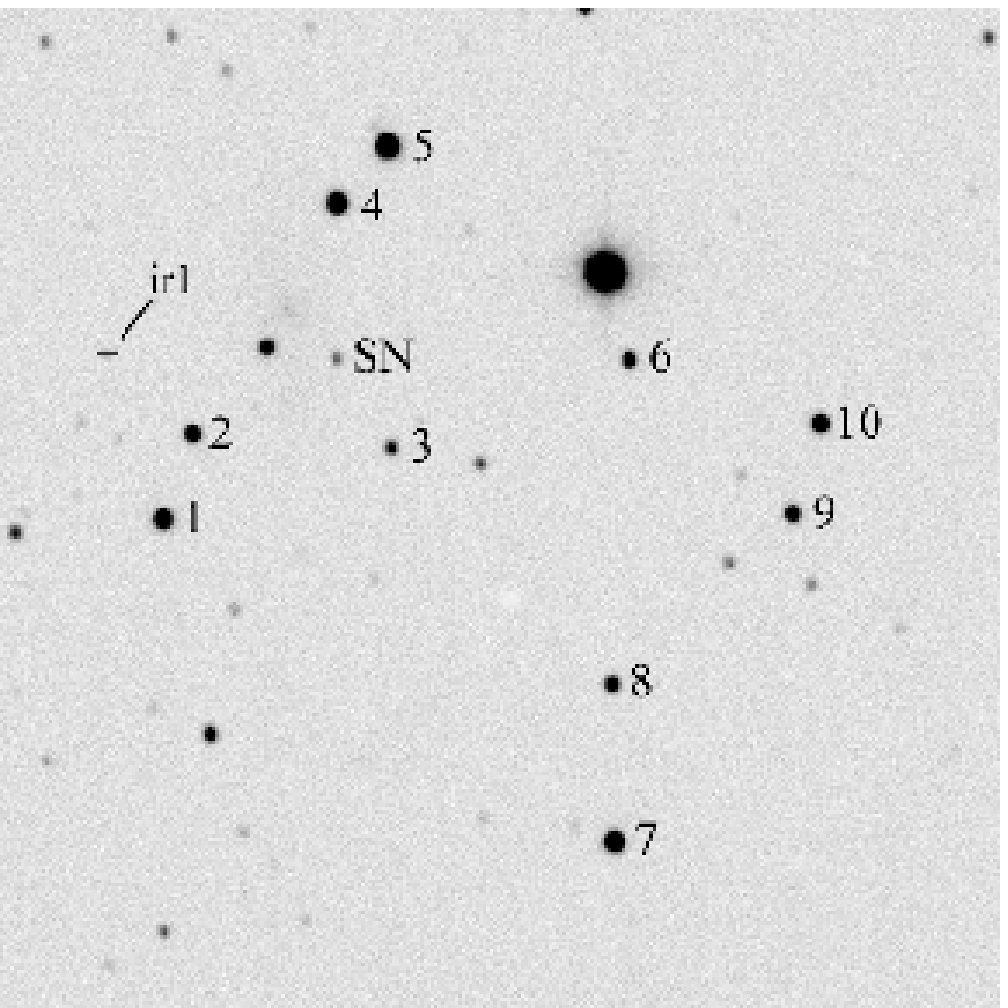, 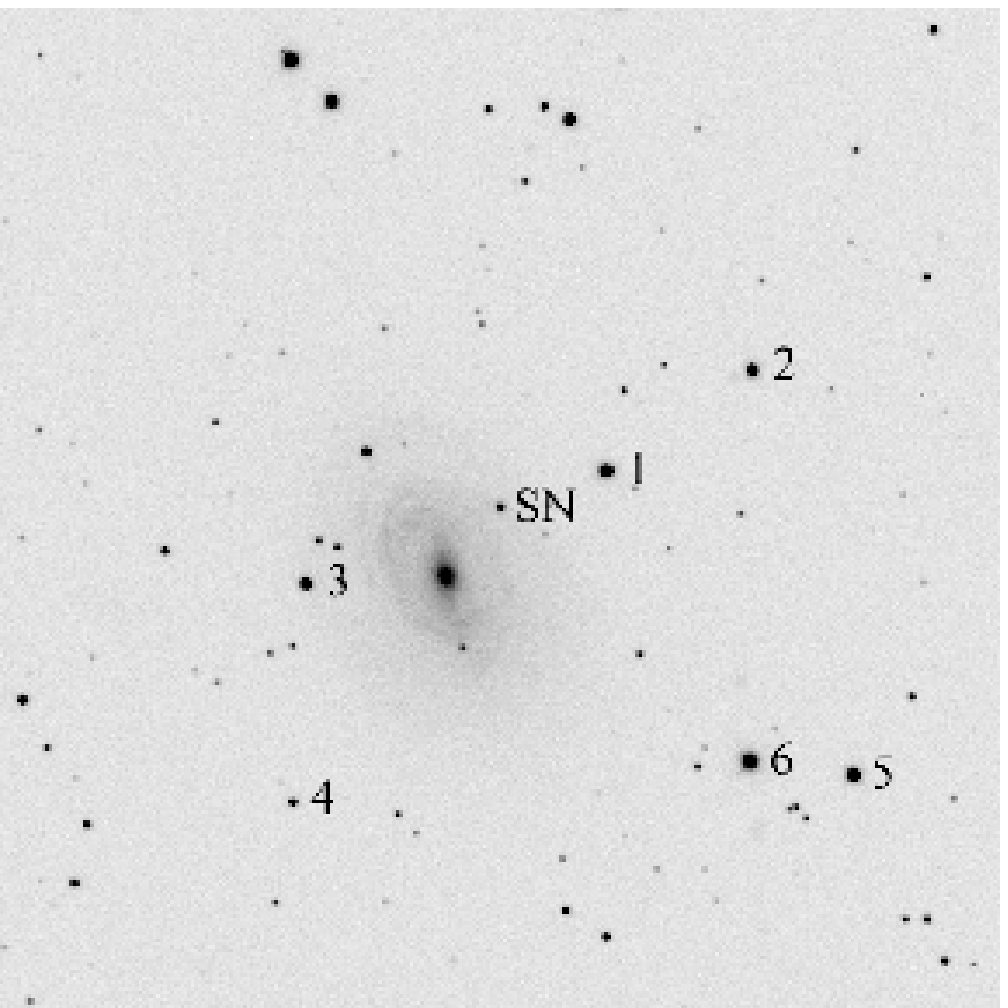, 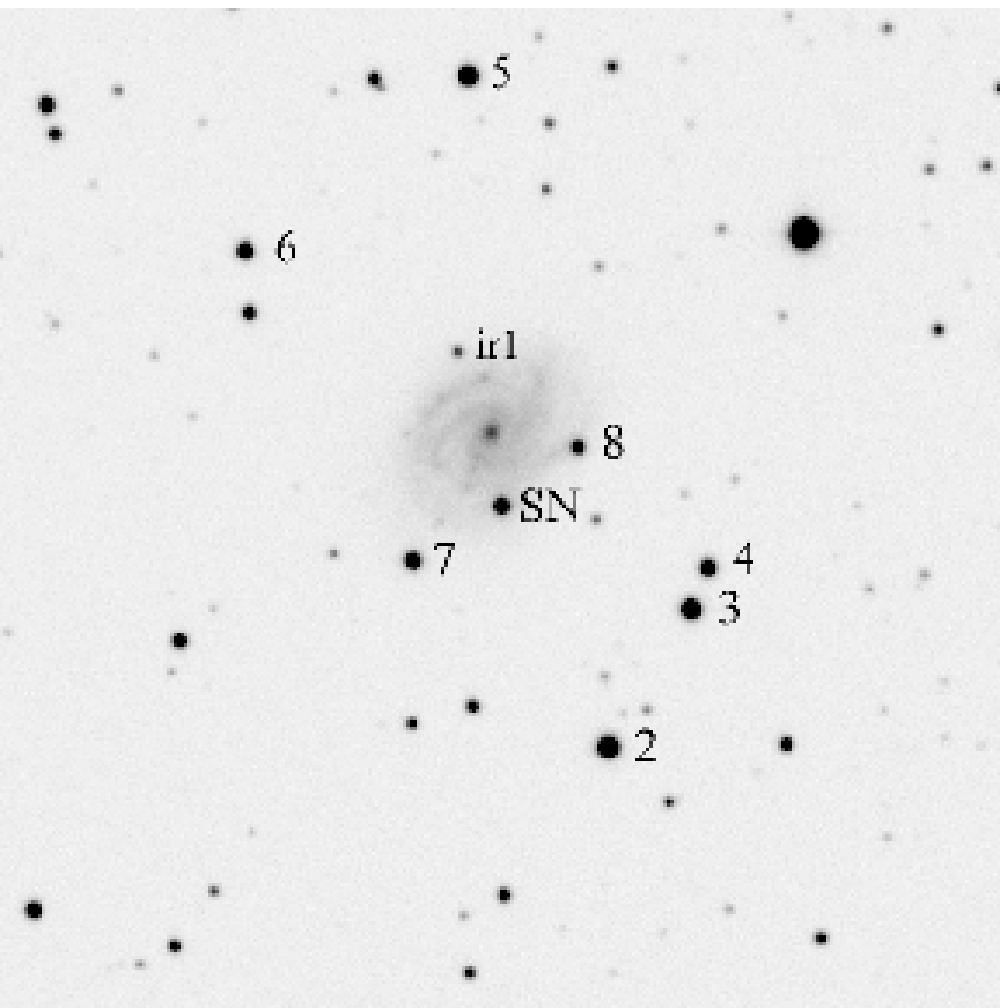, 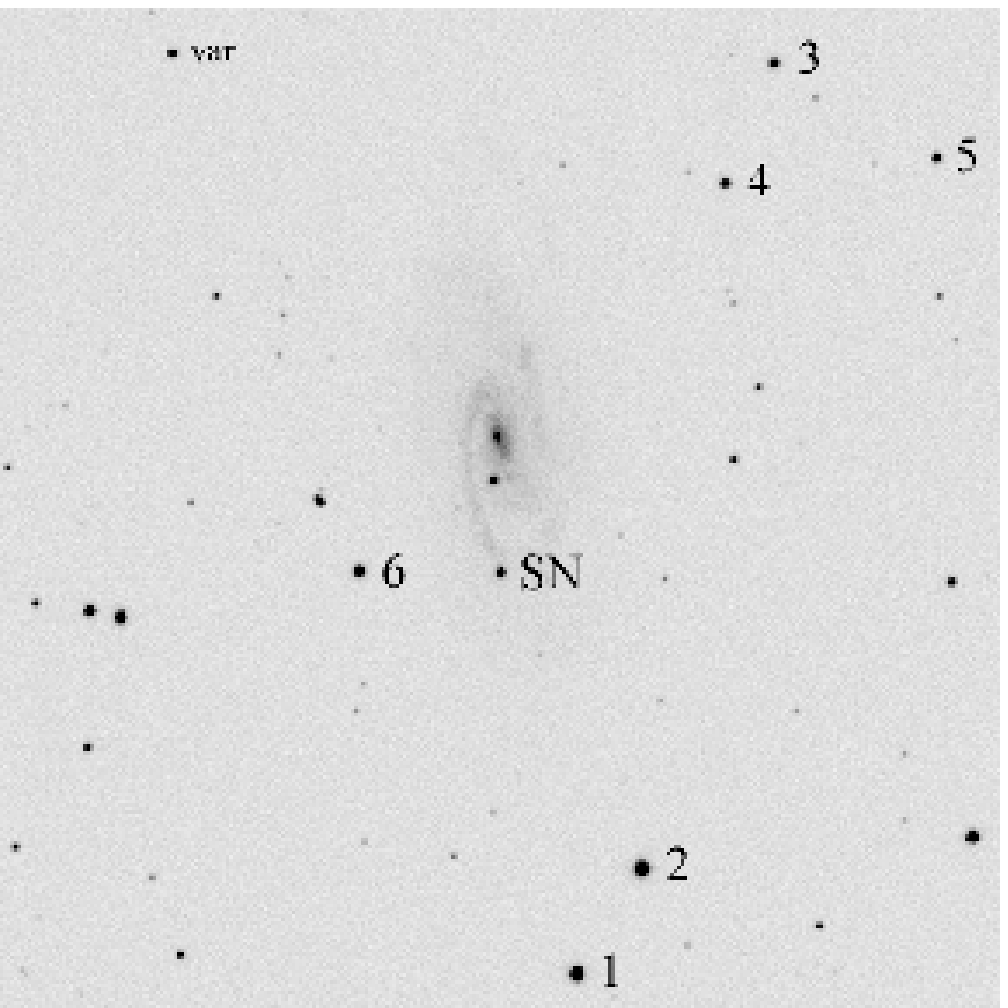,
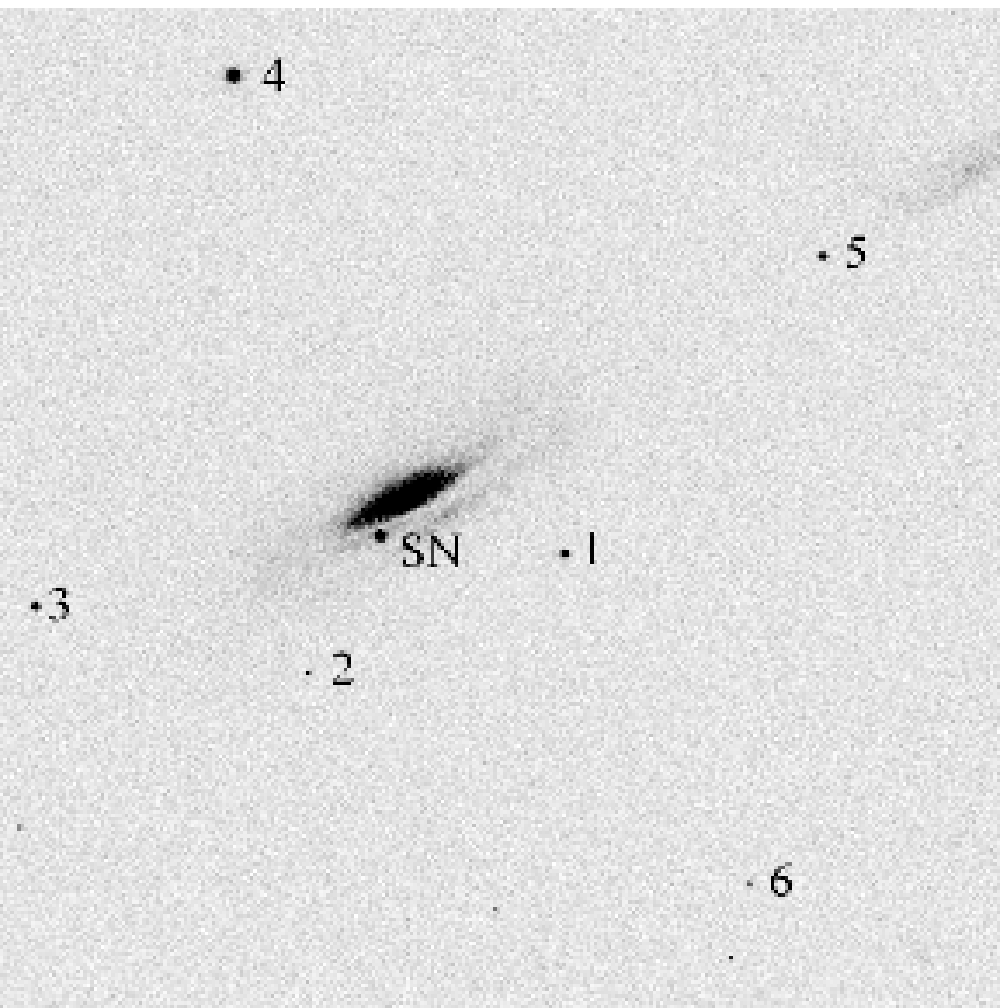] {Finding charts for the supernovae: a) SN~1999ek; 
b) SN~2001bt; c) SN~2001cn; d) SN~2001cz; and e) SN~2002bo.
The object labelled ``var'' in Fig. \ref{find_chart}d is a
previously unknown RR Lyr star; see \citet{Kri_etal04c}.
\label{find_chart}
}

\figcaption[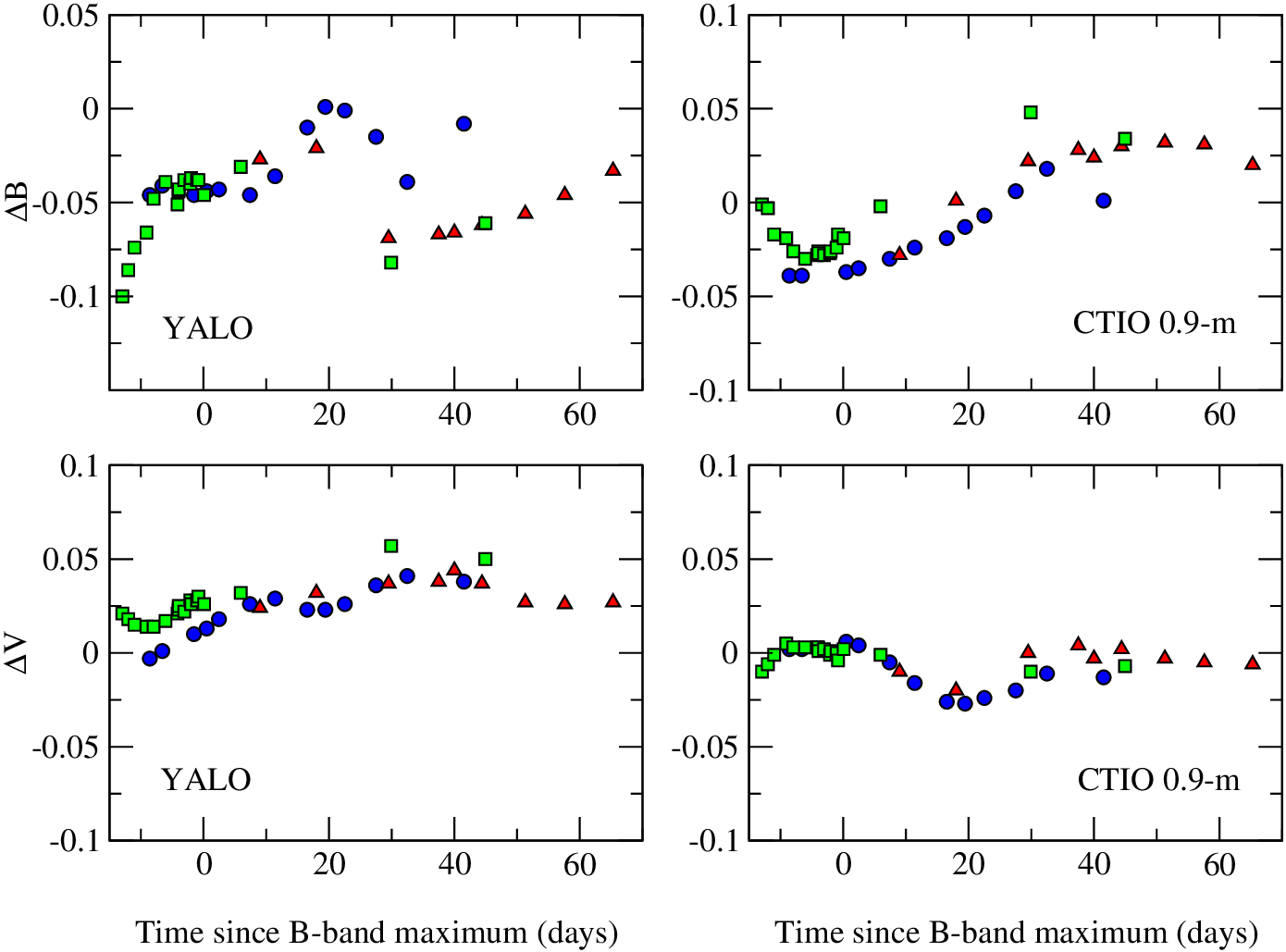] {S-corrections for $B$-band data (top
graphs) and $V$-band data (bottom graphs).  These corrections place
the data from the YALO 1-m and CTIO 0.9-m telescopes on the system of
\citet{Bes90}.  We used spectra of SNe 1999ee (circles), 2001el
(triangles), and 2002bo (squares).  Given the different spectral
energy distributions of these three supernovae, we would not expect
the S-corrections to be identical.
\label{scorr_bv}
}

\figcaption[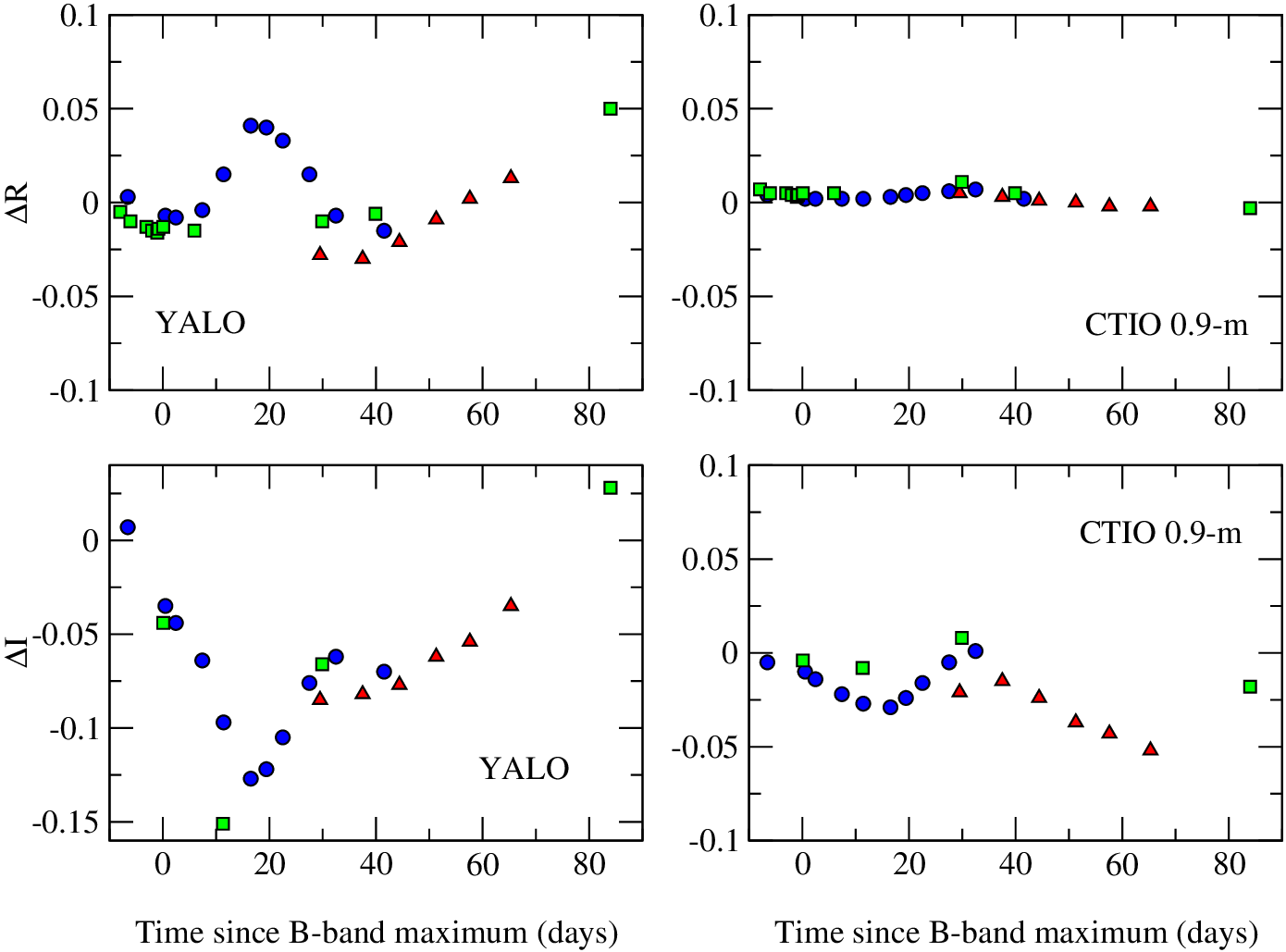] {Similar to Fig. \ref{scorr_bv}, but
S-corrections for R and I.
\label{scorr_ri}
}

\figcaption[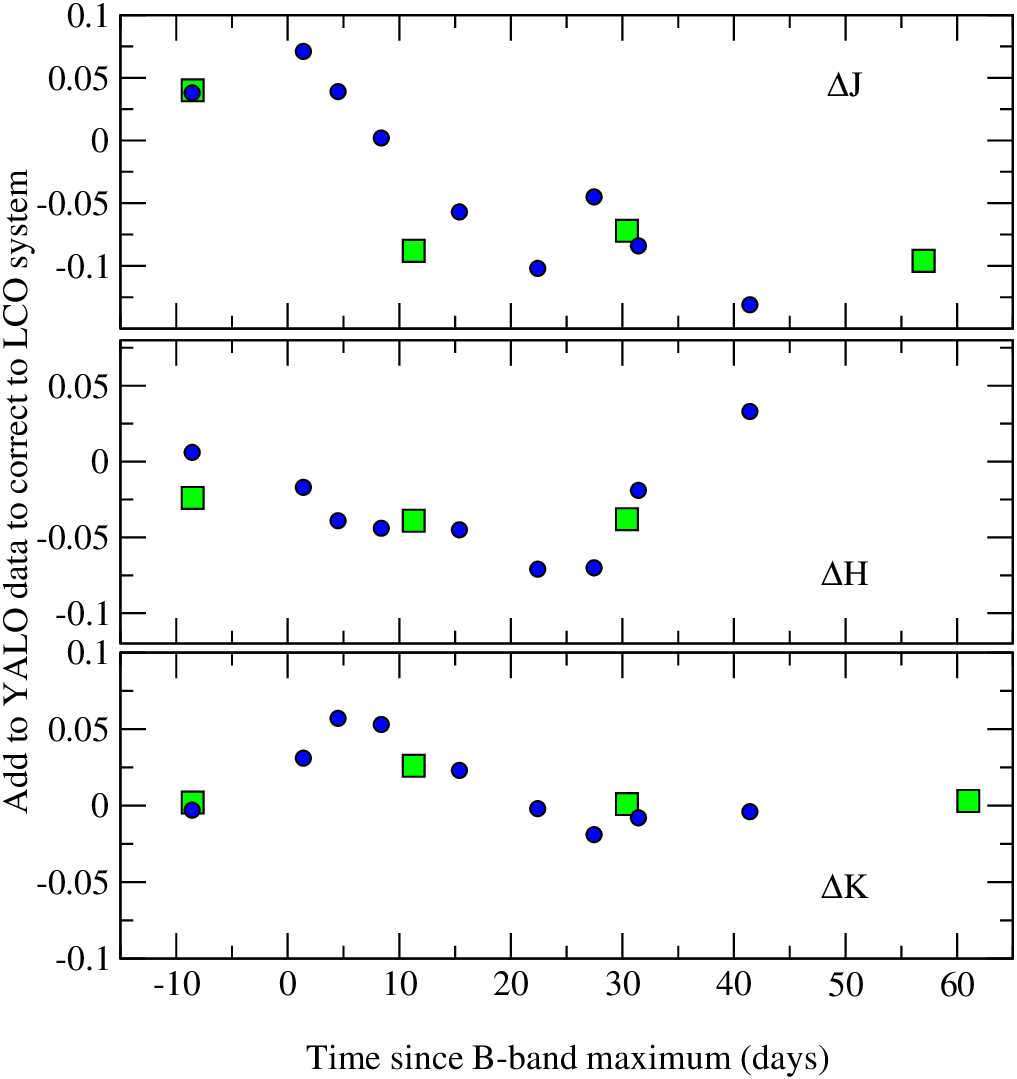] {S-corrections for the infrared bands.
The corrections place the data obtained with the YALO 1-m telescope
on the photometric system of \citet{Per_etal98}.  The filled-in circles
correspond to data derived using (``normal'') spectra of SN~1999ee, while 
the squares correspond to S-corrections obtained from spectra of the
``abnormal'' SN 2002bo.
\label{scorr_jhk}
}

\figcaption[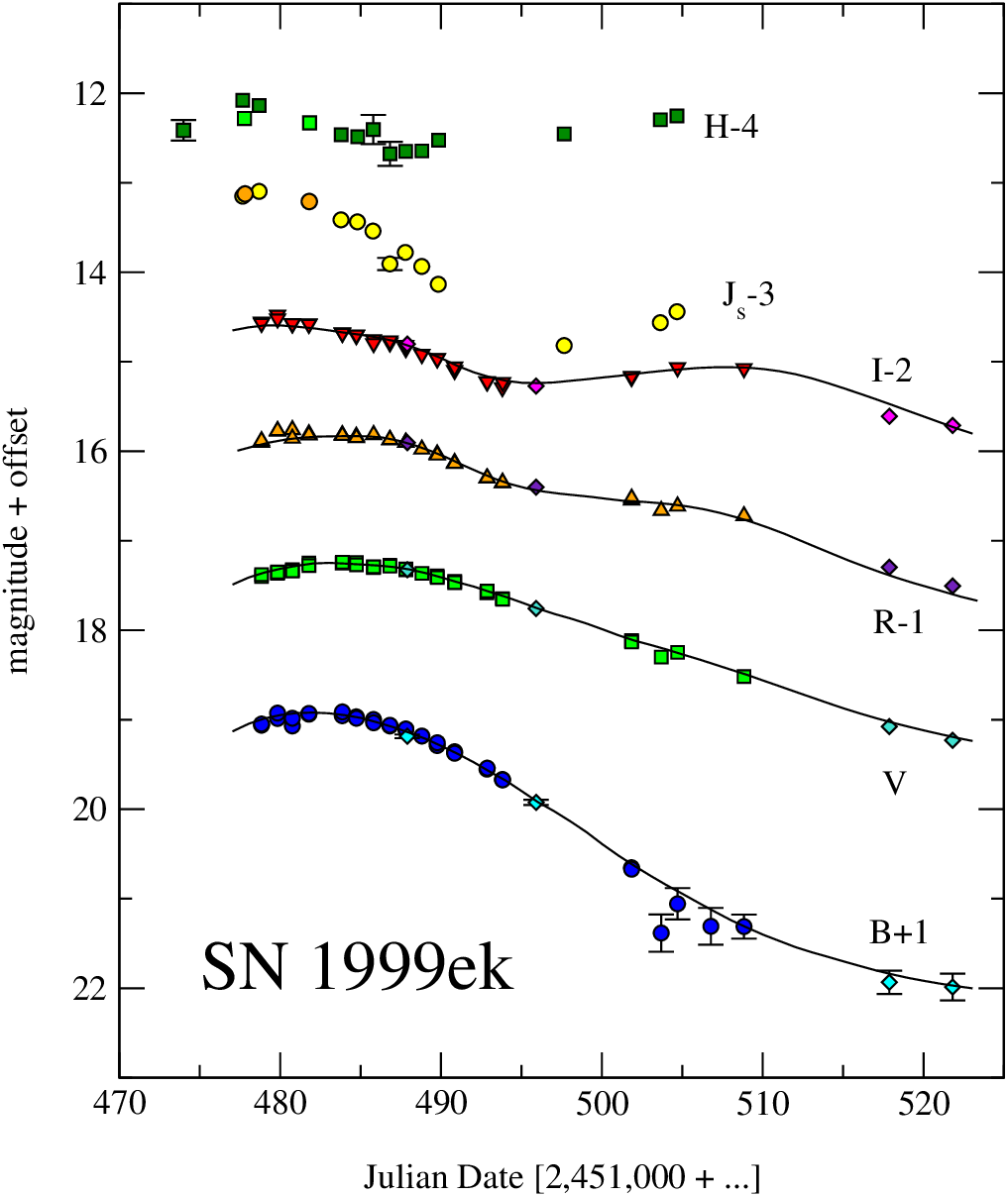] {Optical and IR light curves of SN 1999ek.
Four nights of data from \citet{Jha02} are plotted as diamond-shaped
symbols.
\label{99ek_all}
}

\figcaption[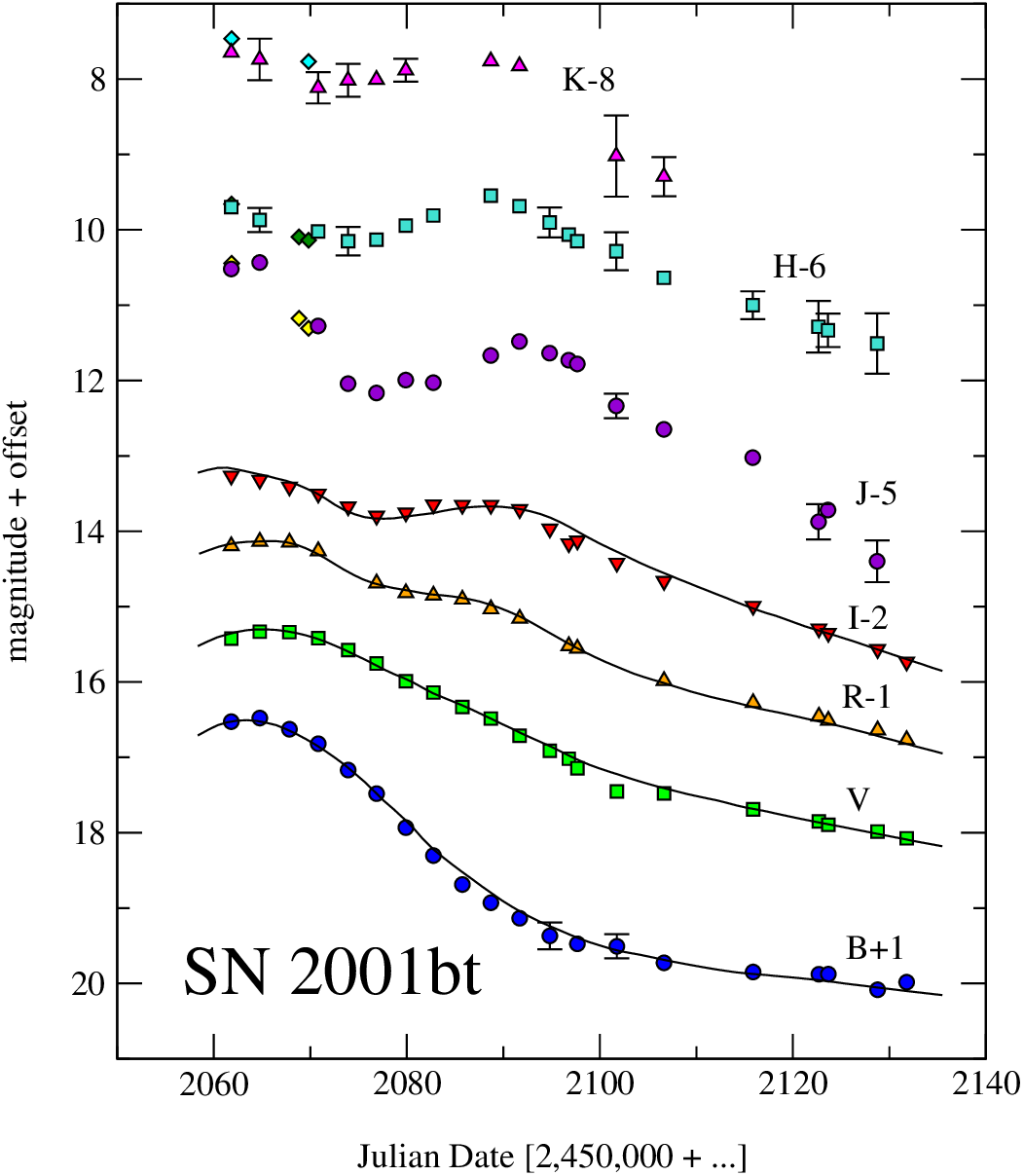] {Optical and IR light curves of SN 2001bt.
Data points with errors greater than 0.15 mag have the error bars
shown.  The infrared data from the LCO 2.5-m telescope, which are
notably more accurate than the data from the YALO 1-m telescope, are
shown as diamond-shaped symbols.  The YALO optical data contain filter
corrections to the filter system of \citet{Bes90}, while the YALO
infrared data are corrected to the system of \citet{Per_etal98}.  The
data are K-corrected, but rescaled in the time axis by 1+$z$ so that
the Julian Dates match the dates of observation.
\label{01bt_all}
}

\figcaption[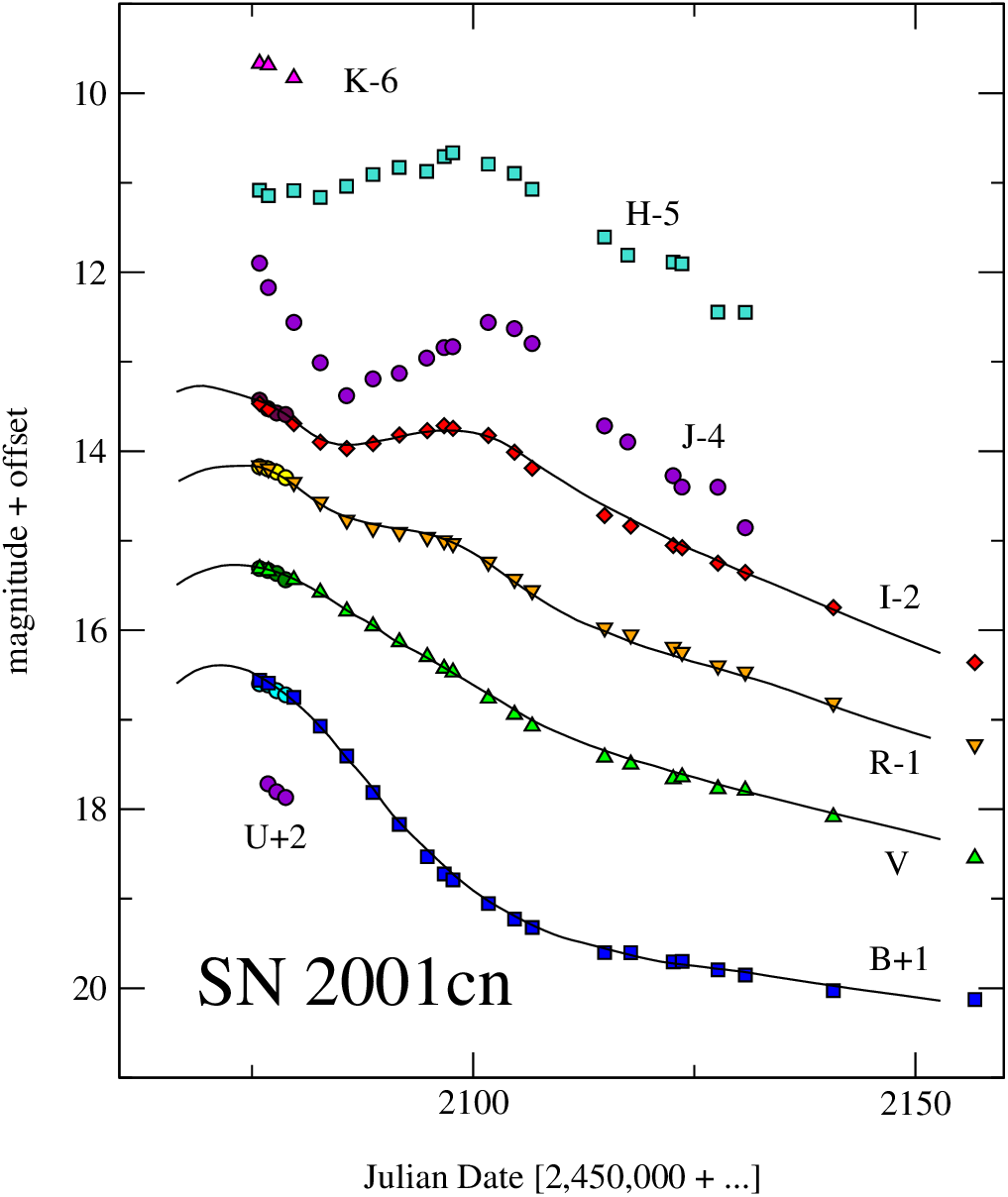] {Optical and infrared light curves of SN 2001cn.
The optical data obtained with the CTIO 0.9-m telescope are shown as filled in
circles.  All the other data were obtained with the YALO 1-m telescope.
As in Fig. \ref{01bt_all} the optical data are S-corrected, K-corrected, and the
Julian Dates are rescaled by 1+$z$ to match the dates of observation.
\label{01cn_all}
}

\figcaption[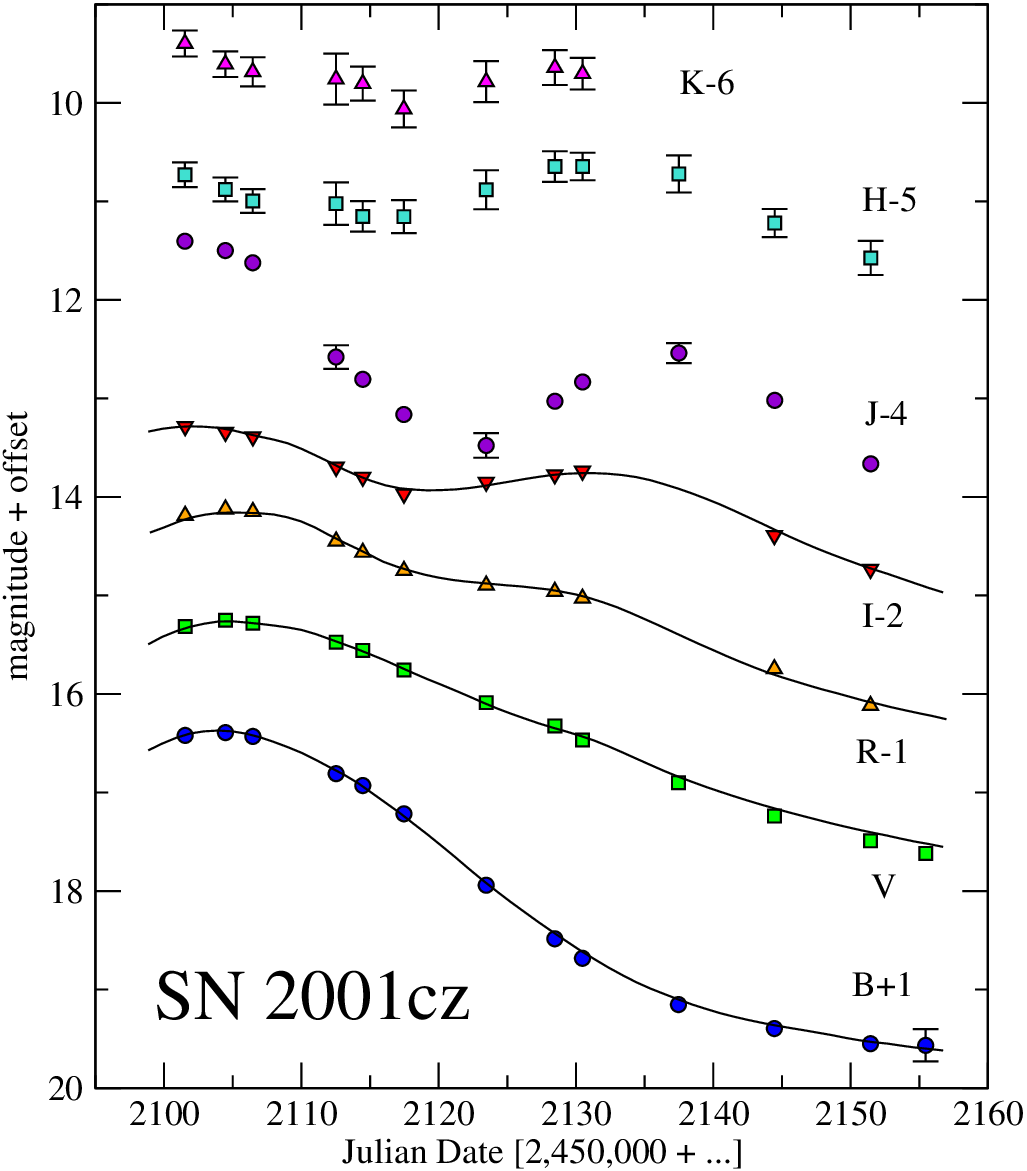] {Optical and infrared light curves of SN 2001cz.
As in Figs. \ref{01bt_all} and \ref{01cn_all} the optical data are S-corrected, 
K-corrected, and the Julian Dates are rescaled by 1+$z$ to match the 
dates of observation.
\label{01cz_all}
}

\figcaption[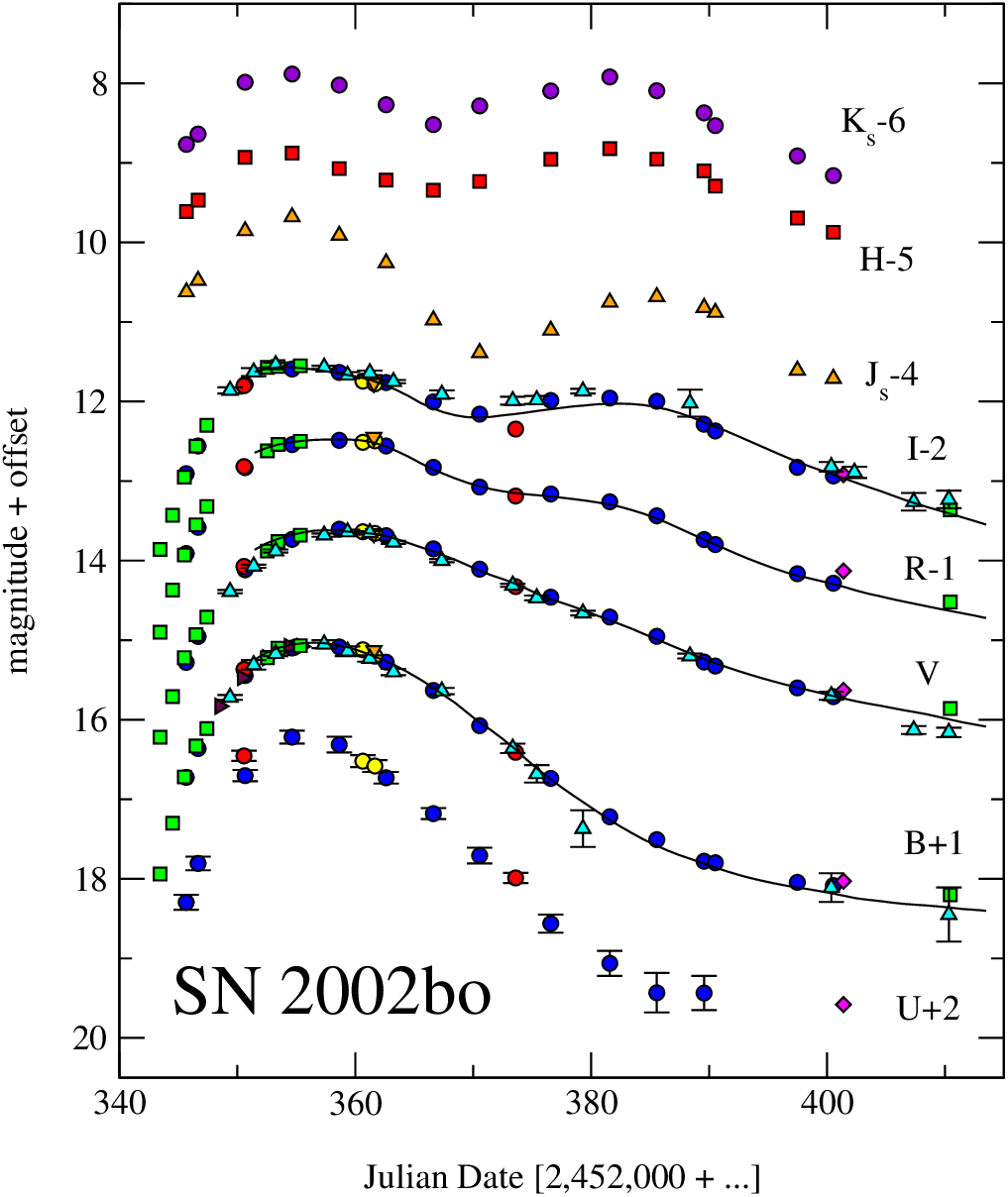] {Optical and IR light curves of SN 2002bo.
The $UBVRI$ data from Cerro Tololo are represented by filled-in
circles, and include filter corrections to the system of
\citet{Bes90}.  The other optical data are from \citet{Ben_etal04}.
The infrared data include the corrections to the photometric system of
\citet{Per_etal98}.
\label{02bo_all}
}

\figcaption[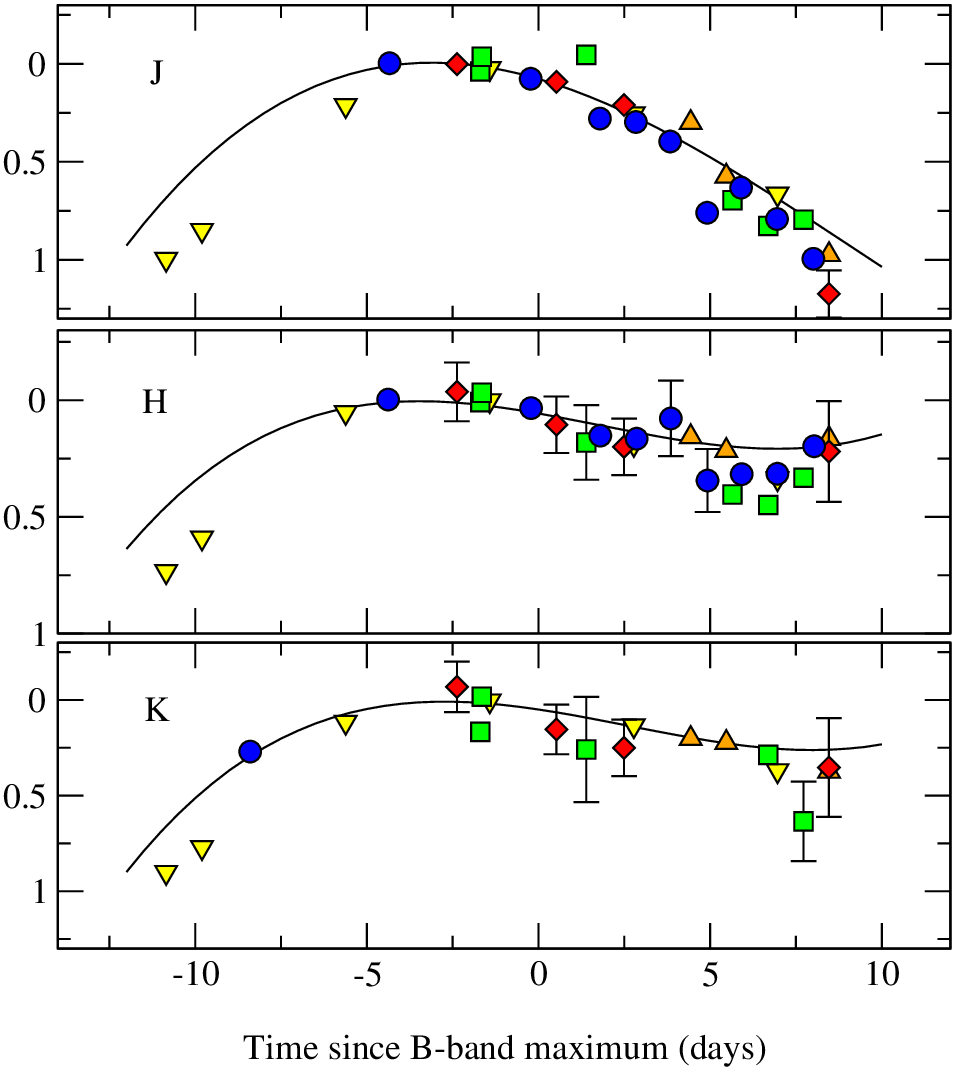] { Photometry of five Type Ia SNe compared to
$JHK$ templates given in Paper V.  Symbols: circles = SN~1999ek;
squares = SN~2001bt; triangles, point up = SN~2001cn; diamonds =
SN~2001cz; and triangles, point down = SN~2002bo.  The time axis
values are stretched according to factors determined from
\dmm.  We show the error
bars if they are greater than $\pm$ 0.10 mag.  For SN~1999ek we do not
plot the data from Steward Observatory.  
\label{jhk_new}
}

\figcaption[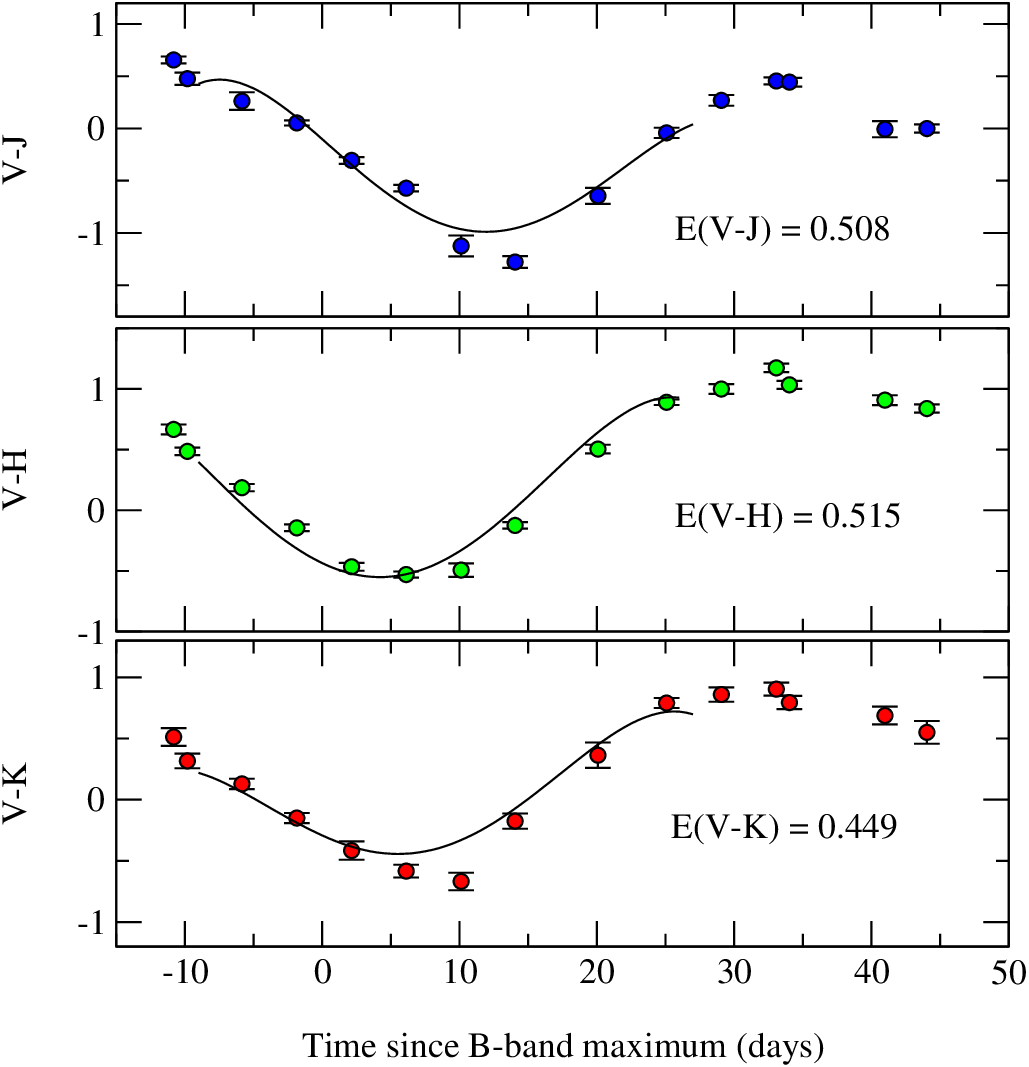] { $V$ {\em minus} near-IR color curves of SN
2002bo.  The fourth order polynomials are the unreddened loci of
mid-range decliners from Paper I, offset by the indicated color
excesses.  These color excesses imply values of A$_V$ considerably
different than one obtains from E($B-V$).  This is further evidence to
the actual spectra presented by \citet{Ben_etal04} that SN~2002bo had
an unusual SED over time.
\label{02bo_vir}
}

\figcaption[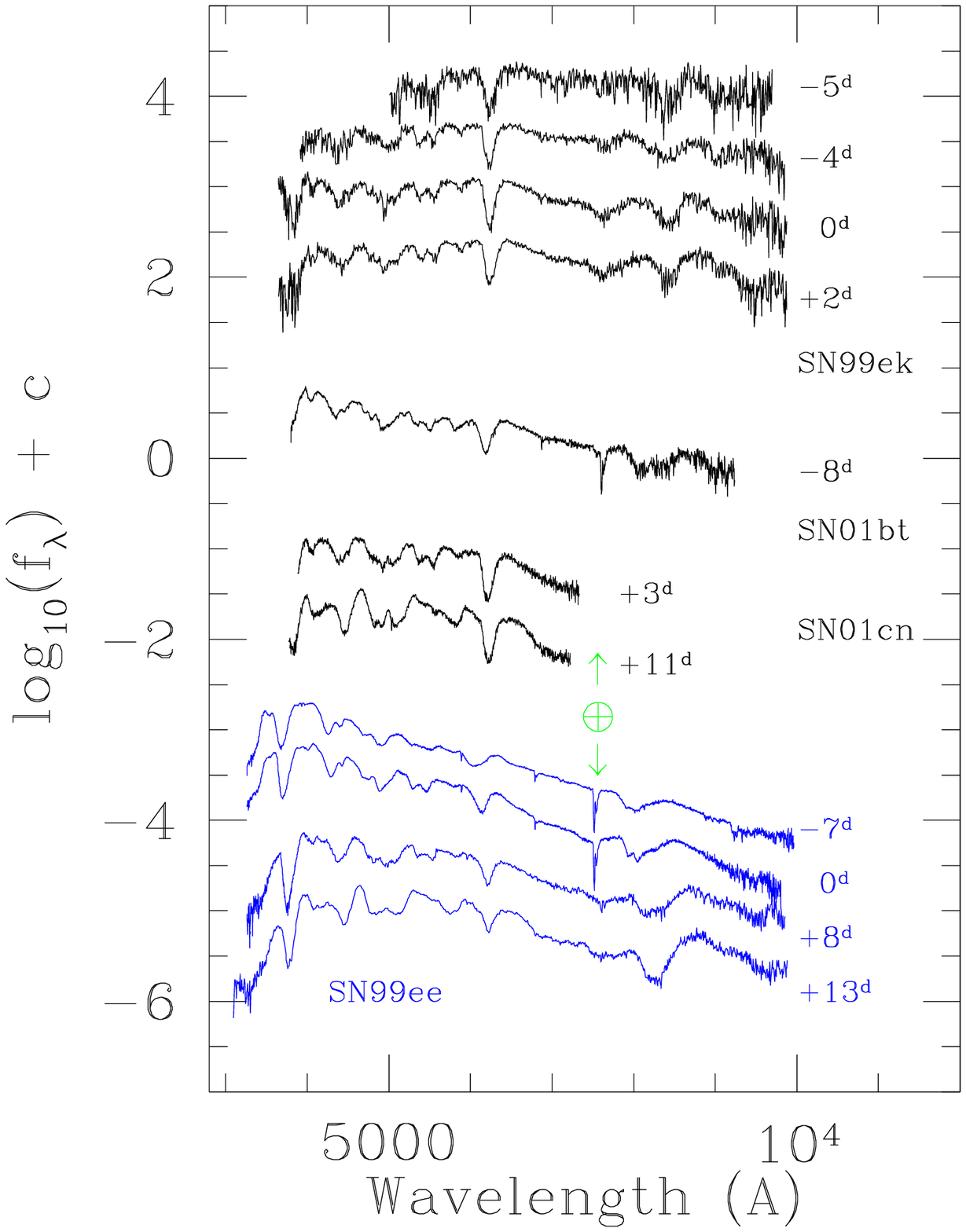] {Spectral atlas of SN~1999ek, SN~2001bt,
SN~2001cn, and SN~1999ee (top to bottom). The top two spectra of
SN~1999ee are taken from \citet{Ham_etal02}. The vertical scale is
given in logarithmic units of ergs s$^{-1}$ cm$^{-2}$ \hspace{1 mm}
\AA$^{-1}$, but includes arbitrary offsets for presentation. The
SN~1999ee spectra cover a range of $-$7 to +13.4 days with respect to
T($B_{max}$).  The absorption feature at $\approx$ 7600 \AA\ in
some of the spectra is the Fraunhofer A band, due to molecular
oxygen in the Earth's atmosphere. 
\label{spectra} 
}

\figcaption[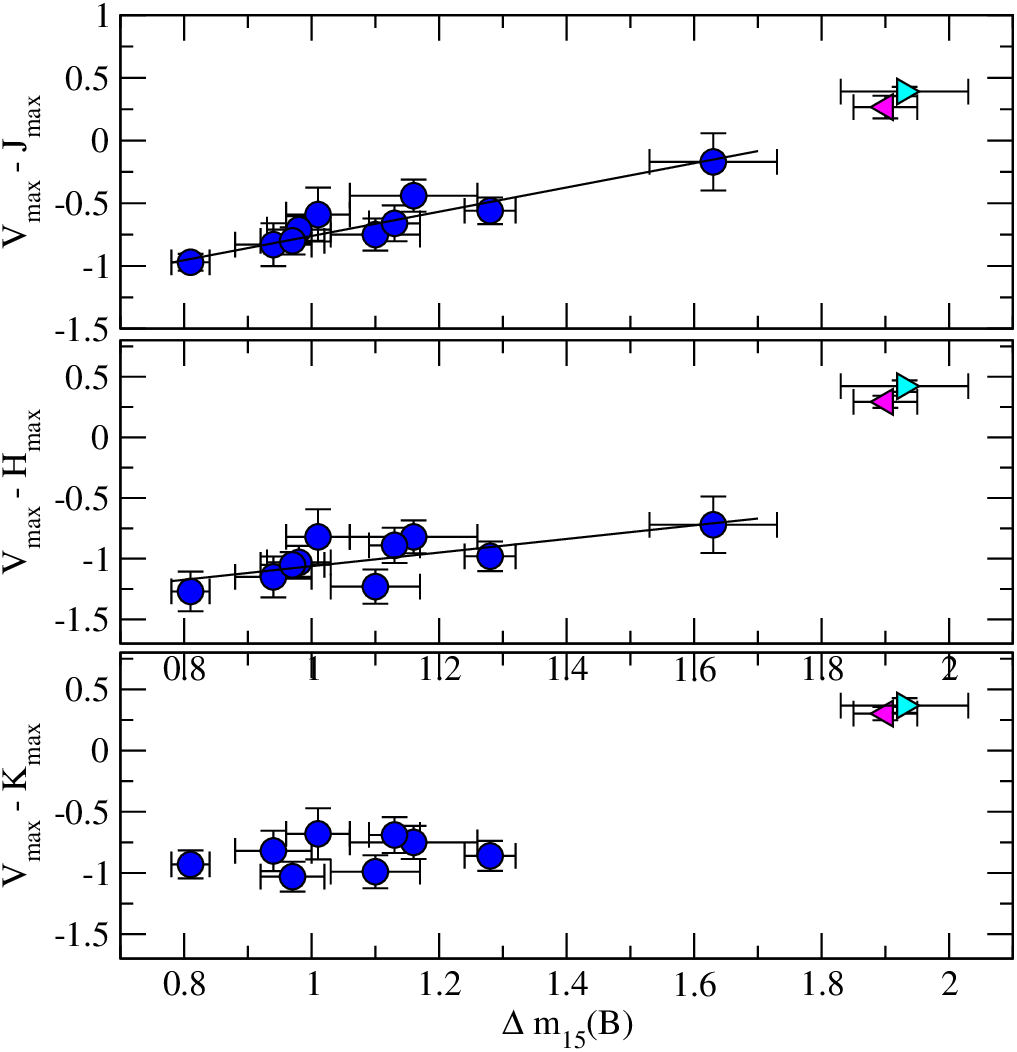] {Dereddened pseudo-colors at maximum light vs.
the decline rate parameter \dmm.  The values of \dmm\ and the
extinction-corrected IR maxima are given in
Tables 15 and 16 of this paper and in Table 1 of \citet{Kri_etal04a}.
The filled-in dots correspond
to the better sampled objects shown in Fig. 12 of \citet{Kri_etal04b}.
The data of SN~1999by (left-pointing triangle in each sub-figure) correspond
to dereddened colors at an epoch 2 days prior to T($B_{max}$) for $V-J$ and 
$V-H$; for $V-K$ the epoch was 3 days prior to T($B_{max}$).
For SN~1991bg (right-pointing triangle in each sub-figure) the data correspond to
an epoch 0.85 days after T($V_{max}$).
\label{vjhkmax} }

\figcaption[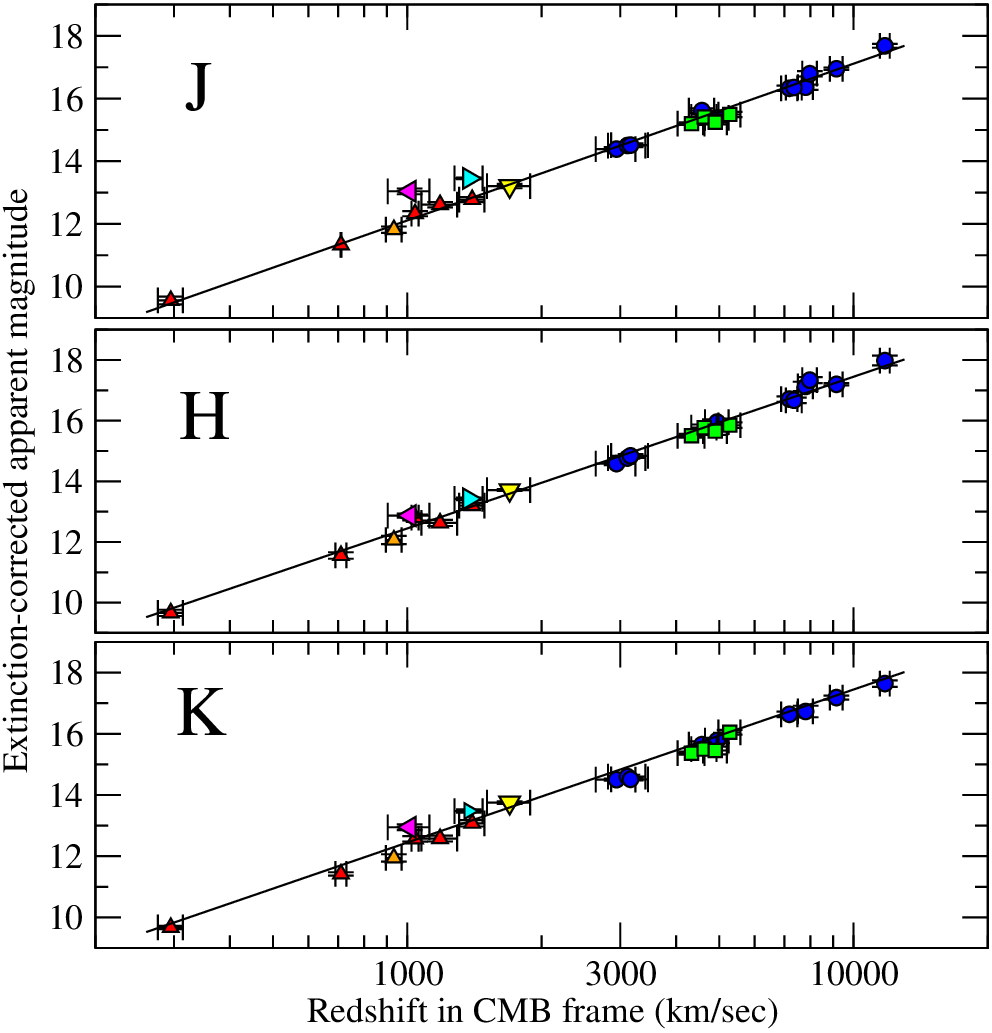] {Hubble diagrams of Type Ia SNe.  We have added
8 objects to Fig. 2 of \citet{Kri_etal04a}.  The triangles
correspond to objects which are not far enough to be in the smooth
Hubble flow.  Their equivalent velocities are determined from direct
measures of the distances to the hosts, on the H$_0$ = 72 km s$^{-1}$
Mpc$^{-1}$ scale of Freedman et al. (2001). SN~2002bo is represented by
downward pointing triangles.  Filled-in circles correspond to objects discussed
by \citet{Kri_etal04a} which are in the smooth Hubble flow.  
Squares correspond to new objects which are in the smooth Hubble flow.
The points at $v_{CMB}$ = 933 km s$^{-1}$ (the Cepheid distance
of NGC 4527 times H$_0$ = 72) correspond to SN~1991T.
The right- and left-pointing triangles correspond to SNe 1991bg and
1999by, respectively, which are subluminous.
\label{jhk_hubble} }

\figcaption[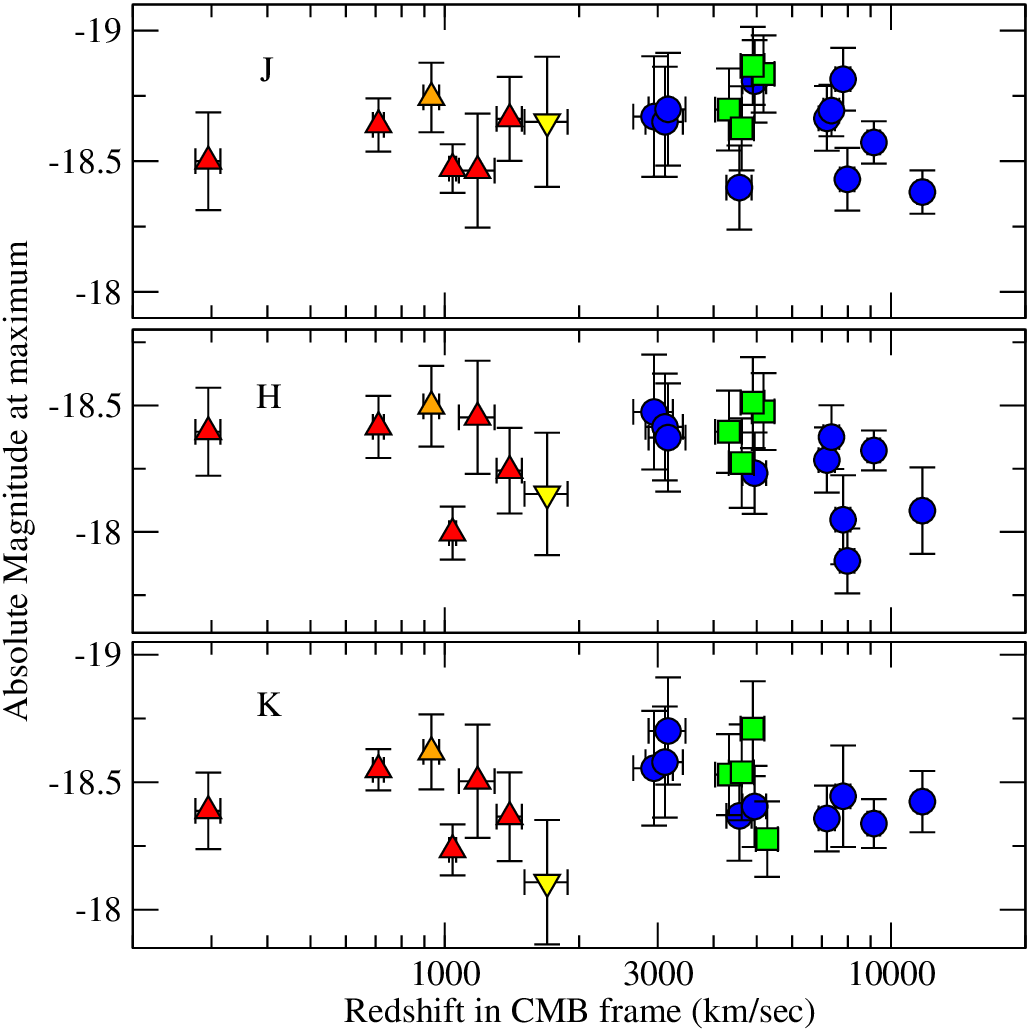] {Absolute magnitudes of Type Ia SNe at
maximum (with \dmm\ in the range 0.8 to 1.7) as a function 
of the logarithm of the redshift.  There is no obvious trend as a function of 
$z$.  The symbols are the same as in Fig. \ref{jhk_hubble}.
\label{absmag_z} }

\figcaption[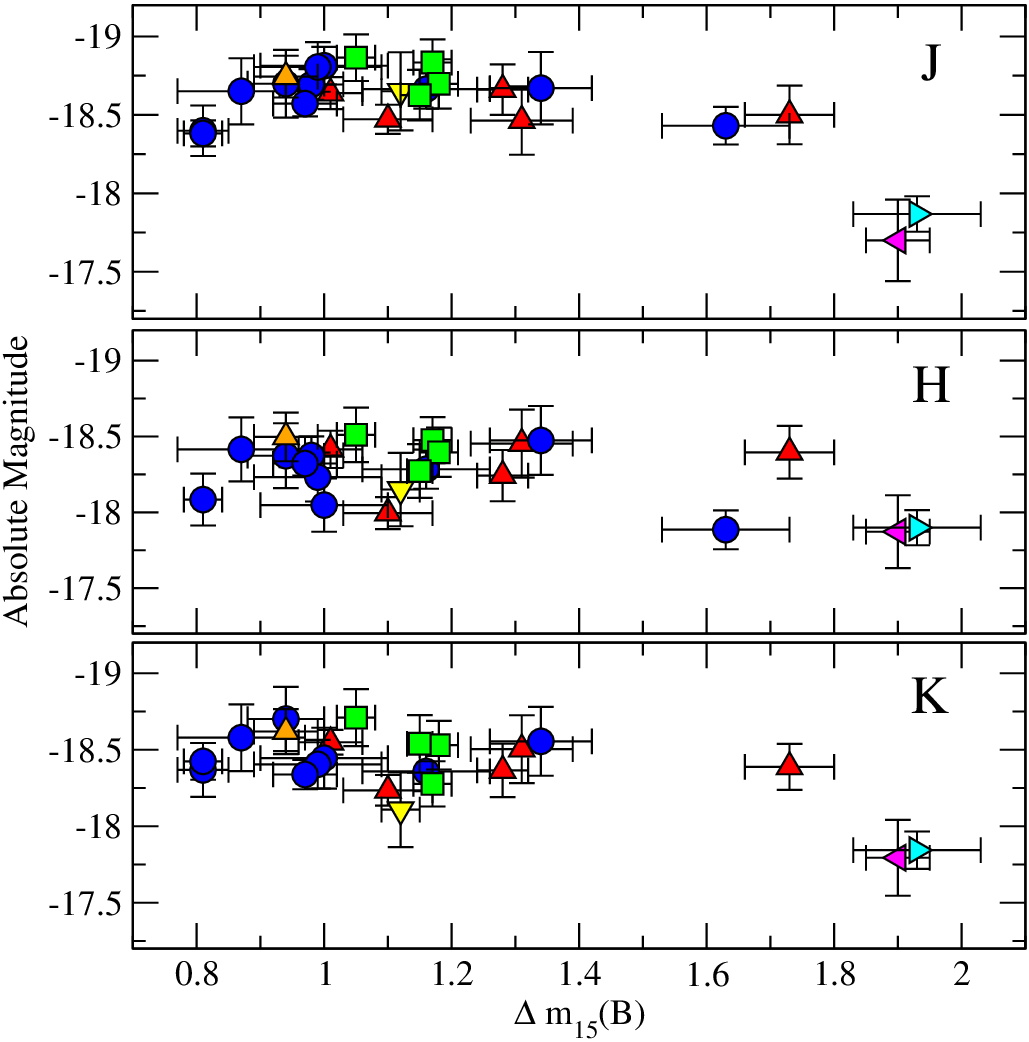] {Absolute magnitudes of Type Ia SNe at
maximum on the H$_0$ = 72 km s$^{-1}$ Mpc$^{-1}$ scale, as a function
of $\Delta$(m)$_{15}$($B$).  We have added 8 objects to Fig. 3 of
\citet{Kri_etal04a}.  The symbols are the same as in
Fig. \ref{jhk_hubble}.
\label{absmag_dm15} }

\clearpage

\begin{figure}
\plotone{SN1999ek.ps}
{\center Krisciunas {\it et al.} Fig. \ref{find_chart}a}
\end{figure}

\begin{figure}
\plotone{SN2001bt.ps}
{\center Krisciunas {\it et al.} Fig. \ref{find_chart}b}
\end{figure}

\begin{figure}
\plotone{SN2001cn.ps}
{\center Krisciunas {\it et al.} Fig. \ref{find_chart}c}
\end{figure}

\begin{figure}
\plotone{SN2001cz.ps}
{\center Krisciunas {\it et al.} Fig. \ref{find_chart}d}
\end{figure}

\begin{figure}
\plotone{SN2002bo.ps}
{\center Krisciunas {\it et al.} Fig. \ref{find_chart}e}
\end{figure}

\begin{figure}
\plotone{scorr_bv.eps}
{\center Krisciunas {\it et al.} Fig. \ref{scorr_bv}}
\end{figure}

\begin{figure}
\plotone{scorr_ri.eps}
{\center Krisciunas {\it et al.} Fig. \ref{scorr_ri}}
\end{figure}

\begin{figure}
\plotone{scorr_jhk.eps}
{\center Krisciunas {\it et al.} Fig. \ref{scorr_jhk}}
\end{figure}

\begin{figure}
\plotone{99ek_all.eps}
{\center Krisciunas {\it et al.} Fig. \ref{99ek_all}}
\end{figure}

\begin{figure}
\plotone{01bt_all.eps}
{\center Krisciunas {\it et al.} Fig. \ref{01bt_all}}
\end{figure}

\begin{figure}
\plotone{01cn_all.eps}
{\center Krisciunas {\it et al.} Fig. \ref{01cn_all}}
\end{figure}

\begin{figure}
\plotone{01cz_all.eps}
{\center Krisciunas {\it et al.} Fig. \ref{01cz_all}}
\end{figure}

\begin{figure}
\plotone{02bo_all.eps}
{\center Krisciunas {\it et al.} Fig. \ref{02bo_all}}
\end{figure}

\begin{figure}
\plotone{jhk_5new.eps}
{\center Krisciunas {\it et al.} Fig. \ref{jhk_new}}
\end{figure}

\begin{figure}
\plotone{02bo_vir.eps}
{\center Krisciunas {\it et al.} Fig. \ref{02bo_vir}}
\end{figure}

\clearpage

\begin{figure}
\plotone{spectra.eps}
{\center Krisciunas {\it et al.} Fig. \ref{spectra}}
\end{figure}

\begin{figure}
\plotone{vjhkmax.eps}
{\center Krisciunas {\it et al.} Fig. \ref{vjhkmax}}
\end{figure}

\begin{figure}
\plotone{hubble12.eps}
{\center Krisciunas {\it et al.} Fig. \ref{jhk_hubble}}
\end{figure}

\begin{figure}
\plotone{absmag_z2.eps}
{\center Krisciunas {\it et al.} Fig. \ref{absmag_z}}
\end{figure}

\begin{figure}
\plotone{absmag9.eps}
{\center Krisciunas {\it et al.} Fig. \ref{absmag_dm15}}
\end{figure}


\newpage


\begin{thebibliography}{}

\bibitem[Alard \& Lupton(1998)]{Ala_Lup98} Alard, C., \& Lupton,
R. H. 1998, \apj, 503, 325

\bibitem[Arce \& Goodman(1999)]{Arc_Goo99} Arce, H. G., \& Goodman,
A. A.  1999, \apjl, 512, L135

\bibitem[Benetti et al.(2002)]{Ben_etal02} Benetti, S., Altavilla, G.,
Pastorello, A., Riello, M., Turatto, M., Cappellaro, E., Tomov, T., \&
Mikolajewski, M.  2002, \iaucirc, 7848

\bibitem[Benetti et al.(2004)]{Ben_etal04} Benetti, S., Meikle, P.,
Stehle, M., et al. 2004, \mnras, 348, 261

\bibitem[Bessell(1990)]{Bes90} Bessell, M. S. 1990, \pasp, 102 1181

\bibitem[Bessell, Castelli, \& Plez(1998)]{Bes_etal98} Bessell, 
M.~S., Castelli, F., \& Plez, B.\ 1998, \aap, 333, 231 

\bibitem[Blackwell et al.(1990)]{Bla_etal90} Blackwell, D.~E., 
Petford, A.~D., Arribas, S., Haddock, D.~J., \& Selby, M.~J.\ 1990, \aap, 
232, 396 

\bibitem[Cacella \& Hirose(2002)]{Cac_Hir02} Cacella, P.. \& Hirose,
Y. 2002, \iaucirc, 7847

\bibitem[Candia et al.(2001)]{Can_etal01} Candia, P., Smith, R. C.,
Suntzeff, N., Norman, D., \& Olsen, K. 2001, \iaucirc, 7644

\bibitem[Candia et al.(2003)]{Can_etal03} Candia, P., Krisciunas, K.,
Suntzeff, N. B., et al. 2003, \pasp, 115, 277 (Paper IV)

\bibitem[Cardelli, Clayton, \& Mathis(1989)]{Car_etal89} Cardelli, J. A., Clayton,
G. C., \& Mathis, J. S. 1989, ApJ, 345, 245

\bibitem[Chassagne (2001a)]{Cha01a} Chassagne, R. 2001a, \iaucirc, 7633

\bibitem[Chassagne (2001b)]{Cha01b} Chassagne, R. 2001b, \iaucirc, 7643

\bibitem[Chassagne (2001c)]{Cha01c} Chassagne, R. 2001c, \iaucirc, 7657

\bibitem[Elias (2003)]{Eli03} Elias, J. H. 2003, private communication

\bibitem[Elias et al.(1981)]{Eli_etal81} Elias, J. H.,
Frogel, J. A., Hackwell, J. A., \& Persson, S. E. 1981, \apj, 251, L13

\bibitem[Elias et al.(1982)]{Eli_etal82} Elias, J. H., Frogel, J. A.,
Matthews, K., \& Neugebauer, G. 1982, \aj, 87 , 1029

\bibitem[Elias et al.(1985)]{Eli_etal85} Elias, J. H., Matthews, G.,
Neugebauer, G., \& Persson, S. E. 1985, \apj, 296, 379

\bibitem[Filippenko et al.(1992)]{Fil_etal92} Filippenko, A. V.,
Richmond, M. W., Branch, D., et al. 1992, \aj, 104, 1543

\bibitem[Filippenko (1997)]{Fil97} Filippenko, A. V. 1997, \araa, 35, 309

\bibitem[Freedman et al.(2001)]{Fre_etal01} Freedman, W. L., Madore,
B. F., Gibson, B. K., et al. 2001, \apj, 553, 47


%
%

\bibitem[Garcia(1993)]{Gar93} Garcia, A. M. 1993, \aaps, 100, 47

\bibitem[Garnavich et al.(2004)]{Gar_etal04} Garnavich, P., Bonanos, A.
Z., Krisciunas, K., et al. 2004, \apj, 613, 1120

\bibitem[Gibson \& Stetson(2001)]{Gib_Ste01} Gibson, B. K., \& Stetson, 
P. B. 2001, \apj, 547, L103

\bibitem[Hamuy et al.(1992)]{Ham_etal92} Hamuy, M., Walker, A. R., 
Suntzeff, N. B., Gigoux, P., Heathcote, S. R., \& Phillips, M. M. 1992, 
\pasp, 104, 533

\bibitem[Hamuy et al.(1993a)]{Ham_etal93a} Hamuy, M., Phillips, M. M.,
Wells, L. A., \& Maza, J. 1993a, \pasp, 105, 787

\bibitem[Hamuy et al.(1993b)]{Ham_etal93b} Hamuy, M., et al.\ 1993b, 
\aj, 106, 2392 

\bibitem[Hamuy et al.(1994)]{Ham_etal94} Hamuy, M., Suntzeff, N. B., 
Heathcote, S. R., Walker, A. R., Gigoux, P., \& Phillips, M. M. 1994, 
\pasp, 106, 566

\bibitem[Hamuy et al.(1996)]{Ham_etal96} Hamuy, M., Phillips, M. M.,
Suntzeff, N. B., Schommer, R. A., Maza, J., Smith, R. C., Lira, P., \&
Aviles, R.  1996, \aj, 112, 2438

\bibitem[Hamuy et al.(2002)]{Ham_etal02} Hamuy, M., Maza, J.
Phillips, M. M., et al. 2002, \aj, 124, 417 

\bibitem[Harrison \& Stringfellow(1991)]{Har_Str91} Harrison, T. E., \& 
Stringfellow, G. 1991, \iaucirc, 5300

\bibitem[H\"{o}flich, Wheeler, \& Thielemann(1998)]{Hoe_etal98} 
H\"{o}flich, P., Wheeler, J.~C., \& Thielemann, F.~K.\ 1998, \apj, 495, 617 

\bibitem[H\"{o}flich et al.(2002)]{Hoe_etal02} H\"{o}flich, P., Gerardy,
C. L., Fesen, R. A., \& Sakai, S. 2002, \apj, 568 , 791

\bibitem[Jha et al.(1999)]{Jha_etal99} Jha, S., Challis, P., Garnavich,
P., Kirshner, R., \& Calkins, M. 1999, \iaucirc, 7300

\bibitem[Jha(2002)]{Jha02} Jha, S. 2002, Harvard University Dissertation

\bibitem[Johnson \& Li(1999)]{Joh_Li99} Johnson, R., \& Li, W. D. 1999,
\iaucirc, 7286

\bibitem[Kawakita et al.(2002)]{Kaw_etal02} Kawakita, H., 
Kinugasa, K., Ayani, K., \& Yamaoka, H. 2002, \iaucirc, 7848

\bibitem[Knop et al.(2003)]{Kno_etal03} Knop, R. A., Aldering,
G., Amanullah, R., et al. 2003, \apj, in press, astro-ph/0309368

\bibitem[Krisciunas et al.(2000)]{Kri_etal00} Krisciunas,
K., Hastings, N. C., Loomis, K., McMillan, R.,
Rest, A., Riess, A. G., \& Stubbs, C. 2000, \apj, 539, 658 (Paper I)

\bibitem[Krisciunas et al.(2001)]{Kri_etal01} Krisciunas, K., Phillips,
M. M., Stubbs, C., et al. 2001, \aj, 122, 1616 (Paper II)

\bibitem[Krisciunas et al.(2003)]{Kri_etal03} Krisciunas, K., Suntzeff, N. B.,
Candia, P., et al. 2003, \aj, 125, 166 (Paper III)

\bibitem[Krisciunas et al.(2004a)]{Kri_etal04a} Krisciunas, K.,
Phillips, M. M., \& Suntzeff, N. B. 2004a, \apjl, 602, L81

\bibitem[Krisciunas et al.(2004b)]{Kri_etal04b} Krisciunas, K., Phillips, M. M.,
Suntzeff, N. B., et al. 2004b, \aj, 127, 1664 (Paper V)

\bibitem[Krisciunas et al.(2004c)]{Kri_etal04c} Krisciunas, K.,
Candia, P., \& Suntzeff, N. B. 2004c, IBVS, No. 5600

\bibitem[Landolt(1992)]{Lan92} Landolt, A. U. 1992, \aj, 104, 340

\bibitem[Layden et al.(1996)]{Lay_etal96} Layden, A. C., Hanson, R. B.,
Hawley, S. L., Klemola, A. R., \& Hanley, C. J. 1996, \aj, 112, 2110

\bibitem[Leibundgut et al.(1993)]{Lei_etal93} Leibundgut, B., Kirshner,
R. P., Phillips, M.  M., et al. 1993, \aj, 105 , 301

\bibitem[Lentz et al.(2000)]{Len_etal00} Lentz, E.~J., Baron, E., 
Branch, D., Hauschildt, P.~H., \& Nugent, P.~E.\ 2000, \apj, 530, 966 

\bibitem[Li et al.(2001a)]{Li_etal01a} Li, W. D., Filippenko, A. V.,
Treffers, R. R., Riess, Adam G., Hu, G., \& Qiu, Y. 2001a, \apj, 546, 734

\bibitem[Li et al.(2001b)]{Li_etal01b} Li, W. D., Filippenko,
A. V., Gates, E., et al. 2001b, \pasp, 113, 1178

\bibitem[Lira et al.(1998)]{Lir_etal98} Lira, P., Suntzeff, N. B., 
Phillips, M. M., et al. 1998, \aj, 115, 234

\bibitem[McGregor (1994)]{McG94} McGregor, P. J. 1994, \pasp, 106, 508

\bibitem[Meikle(2000)]{Mei00} Meikle, W. P. S. 2000, \mnras, 314, 782

\bibitem[Menzies \& Carter(1991)]{Men_Car91} Menzies, J. \& Carter, B. 
1991, \iaucirc, 5246

\bibitem[Nakano et al.(2002)]{Nak_etal02} Nakano, Kushida, Y., \&
Kushda, R.. 2002, \iaucirc, 7847

\bibitem[Nugent et al.(1995)]{Nug_etal95} Nugent, P., Phillips, 
M., Baron, E., Branch, D., \& Hauschildt, P.\ 1995, \apjl, 455, L147 

\bibitem[Pastorello et al.(2001)]{Pas_etal01} Pastorello, A., Altavilla,
G., Benetti, S., Cappellaro, E., \& Turatto, M. 2001, \iaucirc, 7663

\bibitem[Perlmutter et al.(1997)]{Per_etal97} Perlmutter, S., Gabi, S.,
Goldhaber, G., et al. 1997, \apj, 483, 565

\bibitem[Perlmutter et al.(1999)]{Per_etal99} Perlmutter, S., Aldering,
G., Goldhaber, G., et al. 1999, \apj, 517, 565

\bibitem[Persson et al.(1998)]{Per_etal98} Persson, S.~E., Murphy,
D.~C., Krzeminski, W., Roth, M., \& Rieke, M.~J.\ 1998, \aj, 116, 2475

\bibitem[Phillips(1993)]{Phi93} Phillips, M. M. 1993,
\apjl, 413, L105

\bibitem[Phillips et al.(1999)]{Phi_etal99} Phillips, M. M., Lira, P.,
Suntzeff, N. B., Schommer, R. A., Hamuy, M., \& Maza, J. 1999, \aj, 118, 1766

\bibitem[Phillips \& Krisciunas(2001)]{Phi_Kri01} Phillips, M., \&
Krisciunas, K. 2001, \iaucirc, 7633

\bibitem[Porter et al.(1992)]{Por_etal92} Porter, A. C., Dickinson, M.,
Stanford, S. A., Lada, E. A., Fuller, G. A., \& Myers, P. C. 1992, 
\baas, 24 , 1244

\bibitem[Prieto, Rest \& Suntzeff(2005)]{Pri_etal05} Prieto, J. L.,
Rest, A., \& Suntzeff, N. B. 2005, in prepration

\bibitem[Rieke et al.(1993)]{Rie_etal93} Rieke, M.~J., Rieke, 
G.~H., Green, E.~M., Montgomery, E.~F., \& Thompson, C.~L.\ 1993, 
\procspie, 1946, 179 

\bibitem[Riess, Press \& Kirshner(1996)]{Rie_etal96} Riess, A. G.,
Press, W. H., \& Kirshner, R. P. 1996, \apj, 473, 88

\bibitem[Riess et al.(1998)]{Rie_etal98} Riess, A. G., Filippenko, A. V.,
Challis, P., et al. 1998, \aj, 116, 1009

\bibitem[Schlegel, Finkbeiner \& Davis(1998)]{Sch_etal98} Schlegel,
D. J., Finkbeiner, D. P., \& Davis, M. 1998, \apj, 500, 525

\bibitem[Smith(1995)]{Smi95} Smith, H. 1995, RR Lyrae Stars (Cambridge:
Cambridge University Press)

\bibitem[Stanford, Eisenhardt, \& Dickinson(1995)]{Sta_etal95} Stanford, S. A.,
Eisenhardt, P. R. M., \& Dickinson, M. 1995, \apj, 450 , 512

\bibitem[Stetson(1987)]{Ste87} Stetson, P. 1987, \pasp, 99, 191

\bibitem[Stetson(1990)]{Ste90} Stetson, P. 1990, \pasp, 102, 932

\bibitem[Stritzinger et al.(2002)]{Str_etal02} Stritzinger, M., Hamuy,
M., Suntzeff, N. B., et al. 2002, \aj, 124, 2100

\bibitem[Stritzinger \& Leibundgut(2004)]{Str_Lei04} Stritzinger, M.,
\& Leibundgut, B. 2004, \aap, in press, astro-ph/0410686

\bibitem[Strolger et al.(1999)]{Str_etal99} Strolger, L. G., Smith,
R. C., Suntzeff, N. B., Hamuy, M., Phillips, M. M., \& Ugarte,
P. 1999, \iaucirc, 7301

\bibitem[Suntzeff(2000)]{Sun00} Suntzeff, N.~B.\ 2000, 
in AIP Conference Proc. 522, Cosmic Explosions, ed. S. S. Holt \& 
W. W. Zhang (New York: AIP), 65 

\bibitem[Szabo et al.(2003)]{Sza_etal03} Sz\'{a}bo, G. M.,
S\'{a}rneczky, K., Vink\'{o}, J., Cs\'{a}k, B., M\'{e}sz\'{a}ros, S.,
Sz\'{e}kely, P., \& Bebesi, Z. 2003, \aap, 408, 915

\bibitem[Tonry et al.(2000)]{Ton_etal00} Tonry, J. L., Blakeslee,
J. P., Ajhar, E. A., \& Dressler, A. 2000, \apj, 530, 625

\bibitem[Tonry et al.(2001)]{Ton_etal01} Tonry, J. L., Dressler, A.,
Blakeslee, J. P., Ajhar, E. A., Fletcher, A. B., Luppino, G. A.,
Metzger, M. R. \& Moore, C. B. 2001, \apj, 546, 681

\bibitem[Tonry et al.(2003)]{Ton_etal03} Tonry, J. L., Schmidt, B. P.,
Barris, B., et al. 2003, \apj, 594, 1

\bibitem[Wang et al.(2003)]{Wan_etal03} Wang, L., Goldhaber, G.,
Aldering, G., \& Perlmutter, S. 2003, \apj, 590, 944

\end{thebibliography}
\end{document}